\documentclass[aps,a4paper,showpacs,prc,floatfix,twocolumn,amsmath,amssymb]{revtex4-1}
\usepackage{graphicx}
\usepackage{epsfig}
\usepackage{ulem}
\usepackage{morefloats}
\usepackage{rotating}
\usepackage{color}
\usepackage{longtable}

\def\beq{\begin{equation}}
\def\eeq{\end{equation}}
\def\beqa{\begin{eqnarray}}
\def\eeqa{\end{eqnarray}}
\def\ban{\begin{eqnarray*}}
\def\ean{\end{eqnarray*}}
\def\bi{\begin{itemize}}
\def\ei{\end{itemize}}

\begin{document}

\title{The Vlasov formalism for extended relativistic mean field models: the crust-core transition and the stellar matter equation of state}
\author{Helena Pais and Constan{\c c}a Provid\^encia}
\affiliation{CFisUC, Department of Physics, University of Coimbra, 3004-516 Coimbra, Portugal}
\begin{abstract}
The Vlasov formalism is extended to relativistic mean-field hadron models with non-linear terms up to fourth order and applied to the calculation of the  crust-core transition density.
 The effect of the nonlinear $\omega\rho$ and $\sigma\rho$ coupling terms on the crust-core transition density and pressure, and on the macroscopic properties of some families of hadronic stars is investigated. For that purpose, six families of relativistic mean field models are considered.  Within each family, the members differ in the symmetry energy behavior. For all the models, the dynamical spinodals are calculated, and  the crust-core transition density and pressure, and  the  neutron star mass-radius relations are obtained. The effect  on the star radius of the inclusion of a pasta calculation in the inner crust is discussed. The set of six models that best satisfy terrestrial and observational constraints predicts 
a radius of 13.6$\pm$0.3 km  and a crust thickness of $1.36\pm 0.06$km for  a 1.4 $M_\odot$ star.
\end{abstract}

\maketitle
\section{Introduction}
The dynamical response of nuclear matter in the collisionless, low energy regime is adequately described within  the Vlasov equation,  a semi-classical approach that takes into account the correct particle statistics. It is a good approximation to the time-dependent Hartree-Fock equation at low energies \cite{Tang81}, and was used to study heavy-ion collisions at low-intermediate energies \cite{Jin89, Ko88, Ko87, Li88}.
In Refs. \cite{Nielsen-91,Nielsen-93}, the collective modes in cold and hot nuclear matter described with the Walecka model were successfully calculated within this formalism. An  extension of the formalism  was carried out including  non-linear meson terms in the Lagrangian density \cite{Avancini-05,ProvidenciaC-06a, Pais09,Pais10} or  density-dependent meson couplings \cite{magno08}. It has also been shown to be  a good tool to estimate the crust-core transition in cold neutrino-free neutron stars \cite{Avancini-10, Avancini-12}.  Similar calculations  based on the Skyrme interaction had been developed previously \cite{Pethick-95}.   In Refs.  \cite{Avancini-10, Avancini-12}, different calculations of the transition density at several temperatures and isospin asymmetries  were compared, and it  was found that the dynamical spinodal method predicts a lower limit for the crust-core density, and, that the larger the isospin asymmetry, the closer this value is to the Thomas Fermi estimate. For $\beta$-equilibrium matter, i.e., neutron star matter, both results are very similar.

Unified equations of state (EoS), that is,  EoS that describe the neutron star, from its outer crust to the inner core within the same nuclear model, are generally not available. Consequently, the complete EoS is frequently built out of three different pieces, one for the outer crust, another  for the inner crust, and one for the core, obtained from different models. Recently, it was shown in \cite{Fortin16}, that although star properties such as mass and radius do not depend on the  outer crust EoS, the choice of the inner crust EoS  and the matching of the inner crust EoS to the core EoS may be critical, and variations larger than 0.5 km have been obtained for the radius of  a 1.4 $M_\odot$ star. 

For the outer crust, three EoS are used  in the literature: the Baym-Pethick-Sutherland (BPS) EoS \cite{BPS-71}, the Haensel and Pichon (HP) EoS \cite{HP}, or the R\"uster {\it et al} (RHS) EoS \cite{ruester06}. Essentially, the differences existing among them do not affect the mass-radius curves.  This is not anymore true for matter above the neutron drip line. The inner crust of neutron stars may contain exotic geometrical structures termed nuclear pasta \cite{Ravenhall-83} at its upper boarder, just before the crust-core transition.  Several methods have been used to compute these pasta phases: quantum \cite{Watanabe-09}, semiclassical \cite{Horowitz-05,Schneider16} or classical \cite {Dorso-12} Molecular Dynamics calculations, 3D Hartree-Fock calculations \cite{Newton-09,Pais-12} and  Thomas-Fermi (TF) calculations \cite{Maruyama-05,Avancini15, Avancini-10, Avancini-12,okamoto2013}.   The authors of \cite{Pons13}  attribute to the existence of pasta phases  a high impurity parameter corresponding to a large resistivity that causes a very fast magnetic field decay and explains the absence of isolated pulsars with  periods above 12 s.  In the present study, the complete stellar matter EoS will  be constructed by taking (1) a standard EoS for the outer crust, such as BPS, HP or RHS, (2) an adequate inner crust EoS that matches the outer crust EoS at the neutron drip, and the core EoS at the crust-core transition, and (3) the core EoS.  After having the complete EoS,  star properties such as the mass and radius are determined from  the integration of the TOV equations \cite{tolman39, oppenheimer39}.

 Laboratory measurements and first principle calculations put limits on the EoS that describe neutron stars. An example are the microscopic calculations based on nuclear interactions derived from chiral effective field theory \cite{Hebeler-13}, or on realistic two- and three-nucleon interactions, using quantum Monte Carlo (MC) techniques \cite{Gandolfi12}. However, in these approaches there is a reasonable uncertainty associated with the three-body force, meaning that small differences between RMF models and MC results should not be enough to discard those models.
  Other constraints  come from terrestrial experiments, like collective flow data in heavy-ion collisions \cite{Danielewicz-02} and the KaoS experiment \cite{Lynch09}, or the measurement of saturation density properties and properties of nuclei such as the binding energy and rms radii \cite{Tsang-12}. Observational constraints also play a fundamental role. In 2010 and 2013, two very massive pulsars, PSR J1614-2230 \cite{Demorest10}  and PSR J0348+0432 \cite{Antoniadis13}, respectively, both close to $M\sim 2 M_\odot$,  were observed.  An updated determination of the PSR J1614-2230 mass has reduced this value to 1.928$\pm 0.017 M_\odot$ \cite{Fonseca16}. Radii are still not sufficiently constrained, but ESA missions such as the Advanced Telescope for High-energy Astrophysics (Athena+) \cite{athenaplus} and the Theia space mission will allow, amongst other things, to better constraint the $M/R$ relation of neutron stars.

Appropriate nuclear models should  satisfy both kinds of constraints, observational and terrestrial. The NL3  \cite{Lalazissis-97}  parametrization has been fitted to ground-state properties of both stable and unstable nuclei. It is  able to predict 2$M_\odot$, but has a too large symmetry energy slope, and is too hard at high densities. Other  nucleonic EOS that  satisfy the $2M_\odot$ constraint  have been discussed in the literature, see for instance \cite{Lattimer12,Weissenborn12,Bonanno12}.
However,  some nucleonic models agree well with the phenomenology at low and intermediate densities, but fail to produce 2 $M_\odot$ stars because they predict a too soft EoS. Introducing  adequate new non-linear terms in the Lagrangian density will change the density dependence of the EoS,  so as to correct its behavior either by inducing a softening \cite{Horowitz01,Carriere03,tm1} or hardening \cite{Shen-11,Maslov-15} of the EoS. Experimental results at intermediate densities are essential to constrain these terms.

 The transition pressure plays a very important role in the determination of the fraction of the star moment of inertia contained in the crust \cite{Link-99}. The description of glitches considers that the inner crust is a reservoir of angular momentum, and, in Ref. \cite{Link-99}, the authors have estimated that the observed glitches of Vela would be explained if 1.4\% of the total  momentum of inertia of the star resides in the crust. More recently, it was shown that crustal entrainement requires, in fact,  a larger angular momentum reservoir in the crust \cite{Chamel-13}, associated to a fraction of the total momentum of inertia larger than the one the crust may contain. Possible solutions to solve this problem include the contribution of the core to the glitch mechanism \cite{Andersson-12}, or the choice of an appropriate EoS that predicts a sufficiently large pressure  at the crust-core transition \cite{Piekarewicz-14}.

We will analyse the effect of the $\sigma\rho$ and $\omega\rho$ couplings on the crust-core transition density and pressure at zero temperature, starting from three different relativistic mean-field (RMF) models, TM1 \cite{tm1}, NL3 \cite{Lalazissis-97}  and Z271 \cite{Horowitz01,Carriere03}. These three models will be designated  {\it head} of the families of the models that we are going to construct, by adding the terms $\sigma\rho$ or $\omega\rho$ with different coupling strengths. Thus, in the following work, we are going to consider six different RMF families. While NL3 and TM1 have been fitted to the ground state properties of several nuclei, the symmetry energy slope these models predict at saturation density is too high.
Z271 with the NL3 saturation properties has a softer EoS at large densities due to the inclusion of a fourth order term in $\omega$. The first two predict neutron stars with masses above the constraints set by the pulsars J1614-2230 and J0348+0432, while the third fails to satisfy these constraints. However, in Ref. \cite{Dutra-14}, it has been shown that Z271$\omega\rho$5 and Z271$\omega\rho$6 were two of the few models that passed a set of 11 terrestrial constraints. We will consider several strengths of the couplings of the $\sigma\rho$ and $\omega\rho$ terms in order to generate a set of models that span the values of the symmetry energy and its slope at saturation as obtained in different experiments \cite{Tsang-12}. It will be possible to identify  existing correlations between the slope $L$ and the density and pressure transitions in a systematic way, since in each family the isoscalar properties are kept fixed and only the isovector properties change.   We will then select, among the 6 families, the  models that satisfy a set of well accepted saturation properties and constraints from microscopic calculations, and still produce a star with mass larger than $2M_\odot$. Since the Z271 family does not satisfy the last constraint,  we will implement the mechanism proposed in \cite{Maslov-15}, adding an extra non-linear $\sigma$ function that hardens the EoS above saturation.

One of the objectives of the work is to distinguish the effect  of the terms  of the form $\omega\rho$ and $\sigma\rho$ when modifying the density dependence of the symmetry energy. We will consider terms of the form
$f_i^2\rho^2$, with $f_i=\omega$ or $\sigma$, as in Refs. \cite{Horowitz01,Carriere03}, which will affect the neutron skin thickness of nuclei but do not change some well established properties of nuclei, such as the binding energy and the charge radius.  In \cite{Pais16}, where a non-linear $\sigma\rho$ term of the form $\sigma\rho^2$ was used, and a refitting of other parameters kept  the binding energy and the charge radius unchanged,   a non  linear behavior of the transition density  with $L$ was obtained, in contrast to previous studies  \cite{ducoin10,Ducoin11}. We will also pay a special attention to the effect of these terms on the pressure at the crust-core transition, since this is a quantity that directly shows whether the glitch mechanism could be attributed solely to the crust.

In order to calculate the crust core transition properties, we extend the Vlasov formalism previously used to include all non-linear self-interaction and mixing terms involving the $\sigma, \, \omega$ and $\rho$ mesons up to fourth order. Taking the six families of models described above we will determine the crust-core transition density and pressure and  will discuss possible implications of the different density dependence of the symmetry energy on the properties of neutron stars, especially  the effects on the star radius. 

In Section \ref{II}, we present the formalism and derive the dispersion relation within the Vlasov method for the calculation of the dynamical spinodal with the non-linear $\sigma$, $\omega$, and $\rho$ mesons coupling terms. In Section \ref{III}, the formalism is applied to the determination of the crust-core transition, and the construction of a consistent stellar matter EoS, and, finally,  in Sec. \ref{IV}, a few conclusions are drawn.

\section{Formalism} \label{II}

We use the relativistic non-linear Walecka model (NLWM) in the mean-field approximation, within the Vlasov formalism to study nuclear collective modes of asymmetric nuclear matter and  $npe$ matter at zero temperature \cite{Nielsen-91,Nielsen-93}. We will first review the Lagrangian density of the extended RMF model with all meson terms up to quartic order \cite{quarticA, quarticB}, and we will next present the Vlasov formalism to study the collective model of nuclear matter.

\subsection{Extended RMF Lagrangian}
We consider a system of baryons, with mass $M$, interacting with
and through an isoscalar-scalar field $\phi$, with mass $m_s$, an
isoscalar-vector field $V^{\mu}$, with mass $m_v$, and an
isovector-vector field $\mathbf b^{\mu}$, with mass $m_\rho$. When
describing $npe$ matter, we also include a system of electrons with
mass $m_e$. Protons and electrons interact through the
electromagnetic field $A^{\mu}$. The Lagrangian density reads:
$$
{\cal L}=\sum_{i=p,n} {\cal L}_i + {\cal L}_e + {\cal L}_{\sigma} + {\cal L}_{\omega}  + {\cal L}_{\rho} + {\cal L}_{\sigma\omega\rho} + {\cal L}_A
$$
where the nucleon Lagrangian reads
$$
{\cal L}_i=\bar \psi_i\left[\gamma_\mu i D^{\mu}-M^*\right]\psi_i \, ,
$$
with
$$
i D^{\mu}=i\partial^{\mu}-g_v V^{\mu}-
\frac{g_{\rho}}{2}  {\boldsymbol\tau} \cdot \mathbf{b}^\mu - e A^{\mu}
\frac{1+\tau_3}{2} \,,
$$
$$
M^*=M-g_s \phi  \, ,
$$
and the electron Lagrangian is given by
$$
{\cal L}_e=\bar \psi_e\left[\gamma_\mu\left(i\partial^{\mu} + e A^{\mu}\right)
-m_e\right]\psi_e .
$$
The isoscalar part is associated with the scalar sigma ($\sigma$)
field $\phi$, and the vector omega ($\omega$) field $V_{\mu}$,
whereas the isospin dependence comes from  the isovector-vector
rho ($\rho$) field $b_\mu^i$ (where $\mu$ stands for the four
dimensional space-time indices  and $i$ the three-dimensional
isospin direction index). The associated Lagrangians are:
\begin{eqnarray}
{\cal L}_\sigma&=&+\frac{1}{2}\left(\partial_{\mu}\phi\partial^{\mu}\phi
-m_s^2 \phi^2 - \frac{1}{3}\kappa \phi^3 -\frac{1}{12}\lambda\phi^4\right),\nonumber\\
{\cal L}_\omega&=&-\frac{1}{4}\Omega_{\mu\nu}\Omega^{\mu\nu}+\frac{1}{2}
m_v^2 V_{\mu}V^{\mu} + \frac{1}{4!}\xi g_v^4 (V_{\mu}V^{\mu})^2, \nonumber \\
{\cal L}_\rho&=&-\frac{1}{4}\mathbf B_{\mu\nu}\cdot\mathbf B^{\mu\nu}+\frac{1}{2}
m_\rho^2 \mathbf b_{\mu}\cdot \mathbf b^{\mu}, \nonumber\\
{\cal L}_A&=&-\frac{1}{4}F_{\mu\nu}F^{\mu\nu},
\end{eqnarray}
where
$\Omega_{\mu\nu}=\partial_{\mu}V_{\nu}-\partial_{\nu}V_{\mu} ,
\quad \mathbf B_{\mu\nu}=\partial_{\mu}\mathbf b_{\nu}-\partial_{\nu} \mathbf b_{\mu}
- g_\rho (\mathbf b_\mu \times \mathbf b_\nu)$ and $F_{\mu\nu}=\partial_{\mu}A_{\nu}-\partial_{\nu}A_{\mu}$.
We supplement the meson Lagrangian with all the non-linear terms that mix  the $\sigma, \omega$, and $\mathbf{\rho}$ mesons  up to quartic order \cite{quarticA,quarticB,quartic1,quartic2}, 
 \begin{eqnarray}
\label{eq:lnon-lin}
{\cal L_{\sigma\omega\rho}} & =&
\Lambda_{3\sigma}g_{s}g_v^2\phi V_{\mu}V^{\mu}+ 
\Lambda_{2\sigma} g_{s}^2 g_v^2\phi^2V_{\mu}V^{\mu} \nonumber \\
&+&\Lambda_{1\sigma}g_{s}g_{\rho }^{2}\phi \mathbf b_{\mu}\cdot \mathbf b^{\mu} 
+\Lambda_{\sigma} g_s^2 g_\rho^2 \phi^2 \mathbf b_{\mu}\cdot \mathbf b^{\mu} \nonumber \\
&+&\Lambda_{\omega} g_v^2 g_\rho^2 \mathbf b_{\mu}\cdot \mathbf b^{\mu}\, V_{\mu}V^{\mu} .
\end{eqnarray}
An adequate choice of the parameters of the model will allow to build parametrizations compatible with both laboratory measurements and astrophysical observations \cite{quartic1,quartic2}.
The model comprises the following parameters:
three coupling constants, $g_s$, $g_v$, and $g_{\rho}$, of the mesons
to the nucleons, the bare nucleon mass, $M$, the electron mass,
$m_e$, the masses of the mesons, the electromagnetic coupling
constant, $e=\sqrt{4 \pi/137}$, the self-interacting coupling
constants, $\kappa$, $\lambda$, and $\xi$, and the mixing self-interacting coupling constants, $\Lambda_{\omega}, \Lambda_\sigma, \Lambda_{i\sigma}, i=1,2,3$. In this Lagrangian
density, $\boldsymbol \tau$ are the Pauli matrices.

\begin{table*}[t]
\caption{ Symmetric nuclear matter properties at saturation density,
$\rho_0$, for the $\sigma\rho$ models. For TM1, the binding energy per
nucleon, $E/A$, is -16.26 MeV, the incompressibility coefficient, $K$, is 280 MeV, and the nuclear saturation density is 0.145 fm$^{-3}$. For NL3 and Z271, these values are, respectively,  -16.24 MeV, 270/269 MeV, and 0.148 fm$^{-3}$.
 $\rho_{t}$ and $P_{t}$ are the crust-core transition density and pressure, respectively, for $\beta-$equilibrium $pne$ matter, at $T=0$ MeV. $P_N$ is the pressure for neutron matter at $\rho=\rho_0$. The values of the total binding energy per particle ($B/A$), charge radii ($r_c$), neutron radii ($r_n$) and $\Delta r_{\rm np}$, for $^{208}$Pb, are also shown.}  \label{tab1}
  \begin{tabular}{ c ccc cccccccccccccc cccccccc}
    \hline
    \hline
 Model & \phantom{a} & $\Lambda_{\sigma}$ & \phantom{a} & $E_{sym}$ & \phantom{a} & $L$ & \phantom{a} & $K_{sym}$ & \phantom{a} & $K_{\tau}$ & \phantom{a} & $\rho_{t}$ & \phantom{a} & $P_{t}$ & \phantom{a} & $P_N$ & \phantom{a}  &  $B/A$ & \phantom{a} & $r_c$ & \phantom{a} & $r_n$ & \phantom{a} & $\Delta r_{\rm np}$   \\
 & \phantom{a} &  & \phantom{a}  & (MeV)& \phantom{a}  & (MeV)& \phantom{a}  & (MeV)& \phantom{a}  & (MeV)& \phantom{a}  &  (fm$^{-3}$) & \phantom{a} & (MeV/fm$^{3}$) & \phantom{a} & (MeV/fm$^{3}$) & \phantom{a} & (MeV) & \phantom{a} & (fm) & \phantom{a} & (fm) & \phantom{a} & (fm) \\
     \hline
NL3 & \phantom{a} & 0 & \phantom{a} & 37.34 & \phantom{a} & 118 &\phantom{a}& 101 &\phantom{a}& -696 & \phantom{a} & 0.055 & \phantom{a} &  0.258 & \phantom{a} & 5.978 & \phantom{a} &  -7.878 & \phantom{a} & 5.518 &\phantom{a}& 5.740 &\phantom{a}& 0.280 \\

$\sigma\rho$1& \phantom{a}& 0.004 &\phantom{a}&  35.85 &\phantom{a}& 99 &\phantom{a}& 4 &\phantom{a}& -666 &\phantom{a}& 0.058 &\phantom{a}&  0.303 & \phantom{a} & 5.108 &\phantom{a}& -7.891 &\phantom{a}&  5.519 &\phantom{a}& 5.723 &\phantom{a}& 0.262 \\
																								                                                                                     
$\sigma\rho$2 & \phantom{a}& 0.0072 &\phantom{a}& 34.88 &\phantom{a}& 88 &\phantom{a}& -29 &\phantom{a}& -620 &\phantom{a}& 0.063 &\phantom{a}&   0.365  & \phantom{a} & 4.546 &\phantom{a}& -7.897 &\phantom{a}& 5.521 &\phantom{a}& 5.711 &\phantom{a}& 0.249 \\ 
																								                                                                                     
$\sigma\rho$3 &\phantom{a}& 0.011 &\phantom{a}&  33.85 &\phantom{a}& 76 &\phantom{a}& -38 &\phantom{a}& -553 &\phantom{a}& 0.069 &\phantom{a}&  0.408 & \phantom{a} & 4.004 &\phantom{a}& -7.903 &\phantom{a}& 5.523 &\phantom{a}& 5.698 &\phantom{a}&  0.233  \\
																								                                                                                     
$\sigma\rho$4 & \phantom{a} & 0.0145 &\phantom{a}& 33 &\phantom{a}&  68 &\phantom{a}& -28 &\phantom{a}& -487&\phantom{a}& 0.074 &\phantom{a}&  0.468  & \phantom{a} & 3.60 &\phantom{a}& -7.908 &\phantom{a}& 5.526 &\phantom{a}& 5.686 &\phantom{a}& 0.218 \\
																								                                                                                     
$\sigma\rho$5 & \phantom{a}& 0.018 &\phantom{a}&  32.25 &\phantom{a}& 61 &\phantom{a}& -6 &\phantom{a}& -421 &\phantom{a}& 0.078 &\phantom{a}&   0.454   & \phantom{a} & 3.282 &\phantom{a}& -7.911 &\phantom{a}& 5.530  &\phantom{a}& 5.675  &\phantom{a}& 0.203 \\
																								                                                                                     
$\sigma\rho$6 & \phantom{a}& 0.022 &\phantom{a}&  31.47 &\phantom{a}& 55 &\phantom{a}& 24 &\phantom{a}& -348 &\phantom{a}& 0.081 &\phantom{a}&   0.382  & \phantom{a} & 2.991 &\phantom{a}& -7.913 &\phantom{a}& 5.535 &\phantom{a}& 5.662 &\phantom{a}&  0.185 \\

\hline

TM1 & \phantom{a} & 0 &\phantom{a}&  36.84 & \phantom{a} & 111 & \phantom{a} & 34 & \phantom{a}&-517 & \phantom{a} & 0.060 & \phantom{a} & 0.328 & \phantom{a} & 5.479 & \phantom{a} & -7.877 & \phantom{a} & 5.541 & \phantom{a} & 5.753 & \phantom{a}&  0.270 \\

$\sigma\rho$1 & \phantom{a}& 0.004 &\phantom{a} & 34.99 &\phantom{a} & 94 &\phantom{a}&  -29 &\phantom{a}& -499 &\phantom{a}& 0.064 &\phantom{a}& 0.360 & \phantom{a} & 4.724 &\phantom{a}& -7.905 &\phantom{a} & 5.544 &\phantom{a}& 5.737 &\phantom{a}& 0.251 \\   
																								                                                                                            
$\sigma\rho$2 & \phantom{a}& 0.0073 &\phantom{a} & 34.22 &\phantom{a} & 85 &\phantom{a}&  -56 &\phantom{a}& -478 &\phantom{a}& 0.068 &\phantom{a}& 0.407 & \phantom{a} & 4.268  &\phantom{a}& -7.910 &\phantom{a} & 5.545 &\phantom{a}& 5.727 &\phantom{a}& 0.239 \\  
																								                                                                                            
$\sigma\rho$3 & \phantom{a}& 0.011 &\phantom{a}&  33.42 &\phantom{a} & 76 &\phantom{a}&  -67 &\phantom{a}& -443 &\phantom{a}& 0.072 &\phantom{a}&  0.447  & \phantom{a} & 3.829 &\phantom{a}& -7.915 &\phantom{a} & 5.548 &\phantom{a}& 5.716 &\phantom{a}& 0.226  \\  
																								                                                                                            
$\sigma\rho$4 & \phantom{a}& 0.0146 &\phantom{a} & 32.72 &\phantom{a} & 68 &\phantom{a} & -64 &\phantom{a} & -403 &\phantom{a}& 0.076 &\phantom{a}&  0.458 & \phantom{a} & 3.469 &\phantom{a}& -7.918 &\phantom{a} & 5.551 &\phantom{a} & 5.706 &\phantom{a} & 0.213 \\ 
																								                                                                                            
$\sigma\rho$5 & \phantom{a}& 0.019 &\phantom{a}&  31.93 &\phantom{a} & 60 &\phantom{a}&  -50 &\phantom{a}& -350 &\phantom{a}& 0.079 &\phantom{a}&  0.427   & \phantom{a} & 3.104 &\phantom{a}& -7.922 &\phantom{a} & 5.555 &\phantom{a} & 5.694 &\phantom{a} &  0.197 \\
																								                                                                                            
$\sigma\rho$6 & \phantom{a}& 0.022 &\phantom{a}&  31.43 &\phantom{a} & 56 &\phantom{a} & -35 &\phantom{a}  & -313 &\phantom{a}& 0.080 &\phantom{a}&  0.379  & \phantom{a} & 2.896 &\phantom{a}& -7.923 &\phantom{a} & 5.558 &\phantom{a} & 5.686 &\phantom{a}  & 0.186 \\
 
 \hline

Z271 &\phantom{a}& 0 &\phantom{a} &  35.81 &\phantom{a}& 99 &\phantom{a}& -16 &\phantom{a}& -340 &\phantom{a}& 0.068 &\phantom{a}& 0.405 & \phantom{a} & 4.952 & \phantom{a}& -7.777 &\phantom{a}& 5.519 &\phantom{a}& 5.702 &\phantom{a}& 0.241  \\

$\sigma\rho$1 & \phantom{a}& 0.01 &\phantom{a} & 34.87 &\phantom{a}& 87 &\phantom{a}& -65 &\phantom{a}& -350 &\phantom{a}& 0.072 &\phantom{a}&   0.448 & \phantom{a} & 4.370 &\phantom{a}& -7.784 &\phantom{a}& 5.521 &\phantom{a}& 5.691 &\phantom{a}& 0.229 \\  
																								                                                                                         
$\sigma\rho$2 & \phantom{a}& 0.02 &\phantom{a} & 34.00 &\phantom{a}& 77 &\phantom{a}& -92 &\phantom{a}& -344 &\phantom{a}& 0.076 &\phantom{a}&  0.474  & \phantom{a} & 3.850 &\phantom{a}& -7.791 &\phantom{a}& 5.522 &\phantom{a}& 5.680 &\phantom{a}& 0.216 \\  
																								                                                                                         
$\sigma\rho$3 & \phantom{a}& 0.03 &\phantom{a} & 33.21 &\phantom{a}& 68 &\phantom{a}& -104 &\phantom{a}& -327 &\phantom{a}& 0.079 &\phantom{a}&   0.477 & \phantom{a} & 3.399 &\phantom{a}& -7.797 &\phantom{a}& 5.524 &\phantom{a}& 5.670 &\phantom{a}& 0.203 \\   
																								                                                                                         
$\sigma\rho$4 & \phantom{a}& 0.04 &\phantom{a}& 32.47 &\phantom{a}& 60 &\phantom{a}& -106 &\phantom{a}& -304 &\phantom{a}& 0.081 &\phantom{a}&  0.451  & \phantom{a} & 3.015 &\phantom{a}& -7.802 &\phantom{a}& 5.527 &\phantom{a}& 5.659 &\phantom{a}& 0.191 \\  
																								                                                                                         
$\sigma\rho$5 & \phantom{a}& 0.05 &\phantom{a}& 31.79 &\phantom{a}& 53 &\phantom{a}& -100 &\phantom{a}& -276 &\phantom{a}& 0.083  &\phantom{a}&   0.398  & \phantom{a} & 2.691 &\phantom{a}& -7.806 &\phantom{a}& 5.530 &\phantom{a}& 5.649 &\phantom{a}& 0.177 \\   
																								                                                                                         
$\sigma\rho$6 & \phantom{a}& 0.06 &\phantom{a}& 31.15 &\phantom{a}& 48 &\phantom{a}& -88 &\phantom{a}& -245 &\phantom{a}& 0.085 &\phantom{a}&    0.323   & \phantom{a} & 2.418 &\phantom{a}&-7.810  &\phantom{a}& 5.533  &\phantom{a}& 5.639 &\phantom{a}& 0.164 \\  
    \hline
    \hline
  \end{tabular}
\end{table*}

\begin{table*}[t]
\caption{The same as in Table \ref{tab1} for the $\omega\rho$ models.}  \label{tab2}
 \begin{tabular}{cccccccccccccccccccccccccc}
    \hline
    \hline
 Model & \phantom{a} & $\Lambda_{\omega}$ & \phantom{a} & $E_{sym}$ & \phantom{a} & $L$ & \phantom{a} & $K_{sym}$ & \phantom{a} & $K_{\tau}$ & \phantom{a} & $\rho_{t}$ & \phantom{a} & $P_{t}$ & \phantom{a} & $P_N$ & \phantom{a}  &  $B/A$ & \phantom{a} & $r_c$ & \phantom{a} & $r_n$ & \phantom{a} & $\Delta r_{\rm np}$   \\
 & \phantom{a} &  & \phantom{a}  & (MeV)& \phantom{a}  & (MeV)& \phantom{a}  & (MeV)& \phantom{a}  & (MeV)& \phantom{a}  &  (fm$^{-3}$) & \phantom{a} & (MeV/fm$^{3}$) & \phantom{a} & (MeV/fm$^{3}$) & \phantom{a} & (MeV) & \phantom{a} & (fm) & \phantom{a} & (fm) & \phantom{a} & (fm) \\
     \hline
NL3 & \phantom{a} & 0 & \phantom{a} &  37.34 & \phantom{a} & 118 &\phantom{a}& 100 &\phantom{a}& -696 &\phantom{a}& 0.055 &\phantom{a}&  0.258  & \phantom{a} & 5.978 & \phantom{a} &  -7.878 & \phantom{a} & 5.518 &\phantom{a}& 5.740 &\phantom{a}& 0.280 \\

$\omega\rho$1 & \phantom{a} & 0.005 &\phantom{a}& 36.01 &\phantom{a}& 101 &\phantom{a}& 1 &\phantom{a}& -680 &\phantom{a}& 0.057 &\phantom{a}&   0.291 & \phantom{a} & 5.146 &\phantom{a}& -7.891 &\phantom{a}&  5.518 &\phantom{a}& 5.725 &\phantom{a}& 0.265 \\ 
																								                                                                                      
$\omega\rho$2 & \phantom{a} & 0.01 &\phantom{a}& 34.94 &\phantom{a}& 88 &\phantom{a}& -46 &\phantom{a}& -636&\phantom{a}& 0.062 &\phantom{a}&   0.355  & \phantom{a} & 4.511 &\phantom{a}& -7.899 &\phantom{a}& 5.519 &\phantom{a}& 5.712 &\phantom{a}& 0.251  \\ 
																								                                                                                      
$\omega\rho$3 & \phantom{a} & 0.015 &\phantom{a}& 33.98 &\phantom{a}& 77 &\phantom{a}& -60 &\phantom{a}& -578 &\phantom{a}& 0.068 &\phantom{a}&  0.437  & \phantom{a} & 3.998 &\phantom{a}& -7.906 &\phantom{a}& 5.521 &\phantom{a}& 5.700  &\phantom{a}&  0.237 \\ 
																								                                                                                      
$\omega\rho$4 & \phantom{a} & 0.02 &\phantom{a}& 33.12 &\phantom{a}& 68 &\phantom{a}&  -53 &\phantom{a}& -512 &\phantom{a}& 0.074 &\phantom{a}&   0.503  & \phantom{a} & 3.573 & \phantom{a}& -7.912  &\phantom{a}& 5.524 &\phantom{a}& 5.688 &\phantom{a}& 0.223 \\ 
																								                                                                                      
$\omega\rho$5 & \phantom{a} & 0.025 &\phantom{a}& 32.36 &\phantom{a}& 61 &\phantom{a}&  -34 &\phantom{a}& -445 &\phantom{a}& 0.080 &\phantom{a}&   0.533  & \phantom{a} & 3.213 &\phantom{a}& -7.917 &\phantom{a}& 5.526  &\phantom{a}& 5.677  &\phantom{a}& 0.209 \\ 
																								                                                                                      
$\omega\rho$6 & \phantom{a} & 0.03 &\phantom{a}& 31.66 &\phantom{a}& 55 &\phantom{a}& -8 &\phantom{a}& -380 &\phantom{a}& 0.084 &\phantom{a}&    0.516 & \phantom{a} & 2.906 &\phantom{a}& -7.921 &\phantom{a}& 5.530 &\phantom{a}& 5.667 &\phantom{a}&  0.195 \\

\hline

TM1 & \phantom{a} & 0 & \phantom{a} &  36.84 & \phantom{a} & 111 & \phantom{a} & 33 & \phantom{a}&-517 & \phantom{a}& 0.060 & \phantom{a}&  0.328 & \phantom{a} & 5.479 & \phantom{a} & -7.877 & \phantom{a} & 5.541 & \phantom{a} & 5.753 & \phantom{a}&  0.270 \\

$\omega\rho$1 & \phantom{a} & 0.005 &\phantom{a}& 35.12 &\phantom{a} &  95 &\phantom{a}& -34 &\phantom{a}& -511 &\phantom{a}& 0.063 & \phantom{a}&   0.350 & \phantom{a} & 4.760 &\phantom{a}& -7.886 &\phantom{a} & 5.542  &\phantom{a}& 5.741 &\phantom{a}& 0.257  \\ 
																								                                                                                             
$\omega\rho$2 & \phantom{a} & 0.01 &\phantom{a}&  34.29 &\phantom{a} &  85 &\phantom{a}& -73 &\phantom{a}& -496 &\phantom{a}& 0.066 & \phantom{a}&   0.397  & \phantom{a} & 4.265 & \phantom{a}& -7.872 &\phantom{a} & 5.542 &\phantom{a}& 5.733 &\phantom{a}&  0.249 \\   
																								                                                                                             
$\omega\rho$3 & \phantom{a} & 0.015 &\phantom{a}& 33.54 &\phantom{a} &  76 &\phantom{a}& -91 &\phantom{a}& -468 &\phantom{a}& 0.071  & \phantom{a}&   0.455 & \phantom{a} & 3.845 &\phantom{a}& -7.857 &\phantom{a} & 5.543 &\phantom{a}& 5.724 &\phantom{a}& 0.239  \\  
																								                                                                                             
$\omega\rho$4 & \phantom{a} & 0.02 &\phantom{a}& 32.84 &\phantom{a} & 68 &\phantom{a} & -94 &\phantom{a} & -432 &\phantom{a}& 0.075 & \phantom{a}&   0.495 & \phantom{a} & 3.484 &\phantom{a}& -7.828 &\phantom{a} & 5.545 &\phantom{a} & 5.715 &\phantom{a} & 0.228  \\
																								                                                                                             
$\omega\rho$5 & \phantom{a} & 0.025 &\phantom{a}& 32.20 &\phantom{a} & 61 &\phantom{a} & -86 &\phantom{a} & -392 &\phantom{a}& 0.079 & \phantom{a}&   0.510 & \phantom{a} & 3.167 &\phantom{a}& - 7.817 &\phantom{a} & 5.548 &\phantom{a} & 5.703 &\phantom{a} &  0.213 \\ 
																								                                                                                             
$\omega\rho$6 & \phantom{a} & 0.03 &\phantom{a}& 31.61 &\phantom{a} & 56 &\phantom{a} & -72 &\phantom{a}  & -350 &\phantom{a}& 0.082 & \phantom{a}&  0.497  & \phantom{a} & 2.885 &\phantom{a}& -7.791 &\phantom{a} & 5.552 &\phantom{a} & 5.689 &\phantom{a}  & 0.195 \\ 

\hline

Z271 &\phantom{a}& 0 &\phantom{a} & 35.81 &\phantom{a}& 99 &\phantom{a}& -16 &\phantom{a}& -340 &\phantom{a}& 0.068 &\phantom{a}&   0.405 & \phantom{a} & 4.952 &\phantom{a} & -7.777 &\phantom{a}& 5.519 &\phantom{a}& 5.702 &\phantom{a}& 0.241 \\

$\omega\rho$1 & \phantom{a} & 0.01 &\phantom{a}& 35.25 &\phantom{a}& 91 &\phantom{a}& -66 &\phantom{a}& -364 &\phantom{a}& 0.071 &\phantom{a}&  0.436 & \phantom{a} & 4.574 &\phantom{a}& -7.783 &\phantom{a}& 5.519 &\phantom{a}& 5.696 &\phantom{a}& 0.234 \\  
																								                                                                                       
$\omega\rho$2 & \phantom{a} & 0.02 &\phantom{a}& 34.72 &\phantom{a}& 83 &\phantom{a}& -104 &\phantom{a}& -379 &\phantom{a}& 0.072 &\phantom{a}& 0.454 & \phantom{a} & 4.245 &\phantom{a}& -7.788 &\phantom{a}& 5.519 &\phantom{a}& 5.689 &\phantom{a}& 0.228  \\ 
																								                                                                                       
$\omega\rho$3 & \phantom{a} & 0.025 &\phantom{a}& 34.46 &\phantom{a}& 80 &\phantom{a}& -120 &\phantom{a}& -383 &\phantom{a}& 0.073 &\phantom{a}& 0.473 & \phantom{a} & 4.095 &\phantom{a}& -7.791 &\phantom{a}& 5.520 &\phantom{a}& 5.686 &\phantom{a}& 0.225  \\ 
																								                                                                                       
$\omega\rho$4 & \phantom{a} & 0.03 &\phantom{a}& 34.21 &\phantom{a}& 77 &\phantom{a}& -133 &\phantom{a}& -386 &\phantom{a}& 0.074 &\phantom{a}&   0.483  & \phantom{a} & 3.953 &\phantom{a}& -7.793 &\phantom{a}& 5.520 &\phantom{a}& 5.683 &\phantom{a}& 0.222  \\ 
																								                                                                                       
$\omega\rho$5 & \phantom{a} & 0.035 &\phantom{a}& 33.97 &\phantom{a}& 74 &\phantom{a}& -145 &\phantom{a}& -388 &\phantom{a}& 0.075 &\phantom{a}&   0.498  & \phantom{a} & 3.818 &\phantom{a}& -7.796 &\phantom{a}& 5.520 &\phantom{a}& 5.680 &\phantom{a}& 0.218  \\
																								                                                                                       
$\omega\rho$6 & \phantom{a} & 0.04 &\phantom{a}& 33.73 &\phantom{a}&  71 &\phantom{a}& -154 &\phantom{a}& -387 &\phantom{a}& 0.076 &\phantom{a}&   0.509  & \phantom{a} & 3.690 &\phantom{a}& -7.798 &\phantom{a}& 5.520  &\phantom{a}& 5.677 &\phantom{a}& 0.215  \\
																								                                                                                       
$\omega\rho$7 & \phantom{a} & 0.05 &\phantom{a}&  33.27 &\phantom{a}& 65 &\phantom{a}& -168 &\phantom{a}& -383 &\phantom{a}& 0.079 &\phantom{a}&   0.533 & \phantom{a} & 3.451 &\phantom{a}& -7.803 &\phantom{a}& 5.520 &\phantom{a}& 5.671 &\phantom{a}& 0.209  \\  
																								                                                                                       
$\omega\rho$8 & \phantom{a} & 0.06 &\phantom{a}& 32.83 &\phantom{a}& 60 &\phantom{a}&  -178 &\phantom{a}& -376 &\phantom{a}& 0.081 &\phantom{a}&   0.552  & \phantom{a} & 3.230 &\phantom{a}& -7.807 &\phantom{a}& 5.521 &\phantom{a}& 5.665  &\phantom{a}& 0.203  \\
    \hline
    \hline
  \end{tabular}
\end{table*}

\subsection{The Vlasov formalism}

In the sequel we use the formalism developed in Refs. \cite{Nielsen-91,Nielsen-93}, where the collective modes in cold nuclear matter were determined within the Vlasov formalism, based on the Walecka model \cite{Walecka-74}.  In the present section, we extend the Vlasov formalism and include all meson terms up to quartic order. We will use, whenever possible, the  notation introduced in \cite{Nielsen-91,Nielsen-93}.

The time evolution of the distribution functions, $f_i$, is described by the Vlasov equation
\begin{equation}
\frac{\partial f_{i}}{\partial t} +\{f_{i},h_{i}\}=0, \qquad
\; i=p,\,n,\, e,
\label{vlasov1}
\end{equation}
where $\{,\}$ denotes the Poisson brackets. The Vlasov equation  expresses the conservation of the number of particles in phase space, and is, therefore, covariant.

The state that minimizes the energy of asymmetric nuclear matter is characterized
by the Fermi momenta $P_{Fi} ,i = p, n$, $P_{Fe} = P_{Fp}$, and is described by the equilibrium distribution function at zero temperature
\begin{equation}
f_0({\boldsymbol r},{\boldsymbol p})=\mbox{diag}[\Theta(P_{Fp}^2-p^2),\Theta(P_{Fn}^2-p^2),\Theta(P_{Fe}^2-p^2)]
\end{equation}
 and by
the constant mesonic fields, that obey the following equations
\begin{widetext}
\begin{eqnarray}
m_s^2\phi_0 &+& \frac{\kappa}{2} \phi_0^{2} +
\frac{\lambda}{6} \phi_0^{3}- 2\Lambda_\sigma g_\sigma^2 g_\rho^2 \phi_0 b_0^{(0)2}- \Lambda_{1\sigma} g_\sigma g_\rho^2 b_0^{(0)2}- 2\Lambda_{2\sigma} g_\sigma^2 g_v^2 \phi_0 V_0^{(0)2}- \Lambda_{3\sigma} g_\sigma g_v^2 V_0^{(0)2}=g_s\rho_s^{(0)},\,\,\, \nonumber \\
m_v^2\,V_0^{(0)}&+&\frac{1}{6}\xi g_v^4 V_0^{(0)\,3}+2 \Lambda_{\omega}
g_v^2 g_\rho^2 V_0^{(0)} b_0^{(0)\,2}+2 \Lambda_{2\sigma}
g_v^2 g_\sigma^2 V_0^{(0)} \phi_0^2+2 \Lambda_{3\sigma}
g_v^2 g_\sigma V_0^{(0)} \phi_0=g_v j_0^ {(0)},\,\,\, \nonumber \\
m_{\rho}^2\,b_0^{(0)}&+&2\Lambda_{\omega} g_\rho^2 g_v^2  V_0^{(0)\,2}b_0^{(0)}
+2 \Lambda_\sigma g_\rho^2 g_\sigma^2 \phi_0^2 b_0^{(0)}+2\Lambda_{1\sigma} g_\rho^2 g_\sigma \phi_0 b_0^{(0)}=\frac{g_\rho}{2} j_{3,0}^{(0)},\,\,\, \nonumber \\
V^{(0)}_i&=&b_i^{(0)}= A_0^{(0)}= A_i^{(0)}=0,\,\,
\end{eqnarray}
\end{widetext}
where $\rho_s^{(0)}, \, \,  j_0^ {(0)},\, \, j_{3,0}^{(0)}$ are, respectively, the equilibrium scalar density, the nuclear density, and the isospin density.

Collective modes correspond to small oscillations
around the equilibrium state. The linearized equations of motion describe these small oscillations and the collective modes are the solutions of those equations. To construct them, let us define:
\begin{eqnarray*}
f_{i}&=&f_{0 i}^{(0)} + \delta f_{i}, \\ \nonumber
\phi\,&=&\,\phi_0 + \delta\phi, \\ \nonumber
V_0\,&=&\, V_0^{(0)} + \delta V_0 , \quad V_i\,=\,\delta V_i, \\ \nonumber
b_0\,&=&\, b_0^{(0)} + \delta b_0, \quad b_i\,=\,\delta b_i, \\ \nonumber
A_0\,&=&\, \delta A_0, \quad A_i\,=\,\delta A_i.
\end{eqnarray*}

As in \cite{Nielsen-91,Nielsen-93,Avancini-05,Avancini-04}, we
express the fluctuations of the distribution functions in terms of the  generating
functions:
$$S({\mathbf r},{\mathbf p},t)=
\mbox{diag}\left(S_{p},\,S_{n},\,S_{e}\right),$$
such that
$$\delta f_i=\{S_i,f_{0i}\}=-\{S_i,p^2\}\delta(P_{Fi}^2-p^2).$$
The linearized Vlasov equations
for $\delta f_{i}$,
$$\frac{d\delta f_{i}}{d t}+ \{\delta f_{i}, h_{0 i}\}
 +\{f_{0 i},\delta h_{i} \}=0$$
 are equivalent to the following time-evolution equations:

\begin{eqnarray}
\label{eq:deltaf}
\frac{\partial S_{i}}{\partial t} + \{S_{i },h_{0i}\} = \delta h_{i}
&=& -g_s\, \frac{ M^*}{\epsilon_{0}}\delta\phi 
- \frac{{\bf p} \cdot \delta \boldsymbol{\cal
V}_i}{\epsilon_{0}}  + \delta{\cal V}_{0i},\,\, \nonumber \\
i&=&p,n
\end{eqnarray}

\begin{equation}
  \label{eq:deltafe}
\frac{\partial S_{e}}{\partial t} + \{S_{e},h_{0e }\} = \delta h_{e}
= -e\left[ \delta{A}_{0} - \frac{{\bf p} \cdot \delta{\mathbf A}}{\epsilon_{0e}}\right],
\end{equation}
where
\begin{eqnarray*}
\delta{\cal V}_{0i}&=&g_v \delta V_0 + \tau_i \frac{g_{\rho}}{2}\,
\delta b_0 + e\, \frac{1+\tau_{i}}{2}\,\delta A_0 ,\\ \nonumber
\delta \boldsymbol{\cal V}_i&=& g_v \delta {\mathbf V} + \tau_i
 \frac{g_{\rho}}{2} \,\delta {\mathbf b}+ e\, \frac{1+\tau_{i}}{2}\,
\delta {\mathbf A}\\
 h_{0 i}&=&\epsilon_{0} +{\cal V}^{(0)}_{0 i}\,=\sqrt{p^2+M^{*2}}+{\cal V}^{(0)}_{0 i} \\
h_{0 e}&=& \epsilon_{0 e}=\sqrt{p^2+m^{2}_e}.
\end{eqnarray*}
which has only to be satisfied for $p=P_{Fi}$.

The longitudinal modes, with wave vector ${\bf k}$ and frequency $\omega$,
are described by the ansatz
\begin{equation}
\left(\begin{array}{c}
S_{j}({\bf r},{\bf p},t)  \\
\delta\phi  \\
\delta \zeta_0 \\ \delta \zeta_i
\end{array}  \right) =
\left(\begin{array}{c}
{\cal S}_{\omega}^j ({\rm cos}\theta) \\
\delta\phi_\omega \\
\delta \zeta_\omega^0\\ \delta \zeta_\omega^i
\end{array} \right) {\rm e}^{i(\omega t - {\bf k}\cdot
{\bf r})} \;  ,
\label{ansatz}
\end{equation}
where $j=p,\, n,\, e$, $\zeta=V,\, b,\, A$ represent the
vector-meson fields, and $\theta$ is the angle between ${\bf p}$
and ${\bf k}$. The wave vector of the excitation mode,  ${\bf k}$,
is identified with the momentum transferred to the system, that gives rise to the excitation.

For the longitudinal modes,
we get $\delta V_\omega^x = \delta V_\omega^y =0\,$, $\delta b_\omega^x =
\delta b_\omega^y = 0\,$ and $\delta A_\omega^x = \delta A_\omega^y
=0\,$.
Calling $\delta V_\omega^z = \delta V_\omega$, $\delta b_\omega^z = \delta
b_\omega$ and $\delta A_\omega^z = \delta A_\omega$, we will  have for the fields
$ \delta {\cal V}_{i,z}= \delta {\cal V}_\omega^i{\rm e}^{i(\omega t - {\bf k}\cdot
{\bf r})}$ and
$\delta {\cal V}_{0i}= \delta {\cal V}_\omega^{0i}{\rm e}^{i(\omega t - {\bf k}\cdot
{\bf r})}.$
 Replacing the ansatz (\ref{ansatz})
in Eqs. (\ref{eq:deltaf}) and  (\ref{eq:deltafe}), we get

\begin{widetext}
\begin{eqnarray}
i\left(\omega -\omega_{0i}x\right ){\cal S}_{\omega}^i &=& - g_s\frac{M^*}{\epsilon_{F i}}\delta\phi_\omega 
- V_{Fi}x\delta {\cal V}_{ \omega}^{i} \label{Si}\\
i\left(\omega - \omega_{0e}x\right){\cal S}_{\omega}^e &=& -e\delta{A}_\omega^0 +eV_{Fe}x\delta{A}_\omega, \label{Se} \\
\left( \omega^2-k^2-m^2_{s,eff}\right) \delta\phi_\omega &=&-\chi_1\delta b_\omega^0-\chi_2\delta V_\omega^0-\frac{2ig_sM^*}{(2\pi^2)}   
\sum_{i=p,n} P_{Fi}\omega_{0i}\int_{-1}^1 x S_\omega^i(x)dx  \label{fi} \\
\left(\omega^2-k^2-m_{v,eff}^2 \right)\delta V_\omega^0 &=& \chi_v\delta b_\omega^0 +\chi_2 \delta \phi_\omega - \frac{2ig_v}{(2\pi^2)}  
\sum_{i=p,n} \omega_{0i}P_{Fi}\epsilon_{Fi}\int_{-1}^1 x S_\omega^i(x)dx\label{V0} ,\\
\left( \omega^2-k^2-m_{\rho,eff}^2 \right)\delta b_\omega^0 &=&\chi_v\delta V_\omega^0 + \chi_1 \delta \phi_\omega - \frac{ig_{\rho}}{(2\pi^2)} 
\sum_{i=p,n} \tau_i \omega_{0i} P_{Fi}\epsilon_{Fi}\int_{-1}^1 x S_\omega^i(x)dx \label{b0} ,\\
\left( \omega^2-k^2 \right)\delta A_\omega^0 &=&
- \frac{2ei}{(2\pi^2)} \sum_{i=p,e} \omega_{0i}P_{Fi}\epsilon_{Fi} 
\int_{-1}^1 x (S_\omega^p(x)-S_\omega^e(x))dx,
\label{A}
\end{eqnarray}
\end{widetext}
with $x=\cos\theta$, $i=p,n$, $\omega_{0j}=kV_{Fj}=kP_{Fj}/\epsilon_{Fj}$, $j=p,n,e$, $\chi_v=4\Lambda_{\omega}g_v^2g_\rho^2 V_0^{(0)} b_0^{(0)}$, $\chi_1=4\Lambda_\sigma g_\sigma^2 g_\rho^2 \phi_0 b_0^{(0)}+2\Lambda_{1\sigma}g_\sigma g_\rho^2 b_0^{(0)}$, $\chi_2=4\Lambda_{2\sigma} g_\sigma^2 g_v^2 \phi_0 V_0^{(0)}+2\Lambda_{3\sigma}g_\sigma g_v^2 V_0^{(0)}$,
and 
\begin{widetext}
\begin{eqnarray}
m_{s,eff}^2&=&m_s^2+\kappa\phi_0+\lambda/2\phi_0^2-2\Lambda_\sigma g_\sigma^2 g_\rho^2 b_0^{(0)2}-2\Lambda_{2\sigma} g_\sigma^2 g_v^2 V_0^{(0)2}+g_s^2\frac{d\rho_s^0}{dM^*} \nonumber \\
m_{v,eff}^2&=&m_v^2+\frac{1}{2}g_v^4\xi V_0^{(0)2} + 2\Lambda_{\omega} g_v^2 g_\rho^2 b_0^{(0)2}+2\Lambda_{2\sigma} g_\sigma^2 g_v^2 \phi_0^2+2\Lambda_{3\sigma} g_\sigma g_v^2 \phi_0\nonumber \\
m_{\rho,eff}^2&=&m_\rho^2+ 2\Lambda_{\omega} g_\rho^2 g_v^2 V_0^{(0)2}+2\Lambda_{\sigma} g_\sigma^2 g_\rho^2 \phi_0^2+2\Lambda_{1\sigma} g_\sigma g_\rho^2 \phi_0
\end{eqnarray}
\end{widetext}
and from the continuity equation for the density currents, we get for the components of the vector fields
\begin{eqnarray}
k \delta V_\omega &=& \omega \frac{B_v}{B_{v_1}} \delta V_\omega^0 -\frac{\omega}{B_{v_1}}\left(\chi_v\delta b_\omega^0+\chi_2 \delta \phi_\omega  \right)  \label{contV} ,\\
k \delta b_\omega &=& \omega \delta b_\omega^0 -\frac{\omega}{B_\rho}\left(\chi_v \delta V_\omega^0 + \chi_1 \delta \phi_\omega  \right)  \label{contb} ,\\
k \delta A_\omega &=& \omega \delta A_\omega^0  \,.
\label{contA}
\end{eqnarray}
with $B_v=\omega^2-k^2-m_{v,eff}^2$, $B_{v_1}=\omega^2-k^2-m_{v,eff,1}^2$, $B_\rho=\omega^2-k^2-m_{\rho,eff}^2$, and 
$m_{v,eff,1}^2=m_v^2+\frac{1}{6}g_v^4\xi V_0^{(0)2} + 2\Lambda_{\omega} g_v^2 g_\rho^2 b_0^{(0)2}+2\Lambda_{2\sigma} g_\sigma^2 g_v^2 \phi_0^2+2\Lambda_{3\sigma} g_\sigma g_v^2 \phi_0$.

The solutions of Eqs. (\ref{Si})-(\ref{A}) form
a complete set of eigenmodes that may be used to construct a
general solution for an arbitrary longitudinal perturbation. Substituting the set of equations (\ref{fi})-(\ref{A}) into
(\ref{Si}) and (\ref{Se}), we get a set of equations for the
unknowns ${\cal S}_{\omega}^i$, which lead to the following
matrix equation
\begin{equation}
\left(\begin{array}{ccc}
1+F^{pp}L_p&F^{pn}L_p&C_A^{pe}L_p\\
F^{np}L_n&1+F^{nn}L_n&0\\
C_A^{ep}L_e&0&1-C_A^{ee}L_e\\
\end{array}\right)
\left(\begin{array}{c}
A_{\omega p} \\
A_{\omega n} \\
A_{\omega e}
\end{array}\right)=0. \label{matriz}
\end{equation}
with $A_{\omega i}=\int_{-1}^1xS_{\omega i}(x)dx$, $L_i=L(s_i)=2-s_i\ln((s_i+1)/(s_i-1))$, where $s_i=\omega/\omega_{0i}$, and $F^{ij}=C_s^{ij}-C_v^{ij}-C_{\rho}^{ij}-C_A^{ij}\delta_{ip}\delta_{ij}$, and $$C_A^{ij}=-\frac{e^2}{2\pi^2}\frac{1}{k^2}\frac{P_F^{j2}}{V_F^i} \nonumber$$
The coefficients $C_s^{ij}, C_v^{ij}$ and $C_{\rho}^{ij}$ are given by:
\begin{widetext}
\begin{eqnarray}
C_s^{ij}&=&\left(1-\frac{\omega^2}{k^2}\left(\frac{g_v\chi_2}{B_{v_1}}+\frac{g_\rho\tau_i\chi_1}{2B_\rho}\right)\right)\frac{g_\sigma M^*}{2\pi^2 P_{F_i}}\left(f_\sigma^\sigma g_\sigma M^*P_{F_j}V_{F_j}+f_b^\sigma\frac{g_\rho}{2}\tau_j P_{F_j}^2+f_\omega^\sigma g_v P_{F_j}^2 \right) \\
C_v^{ij}&=&\left(1-\frac{\omega^2}{k^2}\left(\frac{B_v}{B_{v_1}}-\frac{g_\rho\tau_i\chi_v}{2g_v B_\rho}  \right)\right)\frac{g_v}{2\pi^2 V_{F_i}}\left(f_\sigma^\omega g_\sigma M^*P_{F_j}V_{F_j}+f_b^\omega\frac{g_\rho}{2}\tau_j P_{F_j}^2+f_\omega^\omega g_v P_{F_j}^2 \right) \\
C_{\rho}^{ij}&=&\left(1-\frac{\omega^2}{k^2}\left(1-\frac{2g_v\chi_v}{g_\rho\tau_i B_{v_1}}  \right)\right)\frac{g_\rho \tau_i}{4\pi^2 V_{F_i}}\left(f_\sigma^b g_\sigma M^*P_{F_j}V_{F_j}+f_b^b\frac{g_\rho}{2}\tau_j P_{F_j}^2+f_\omega^b g_v P_{F_j}^2 \right)
\end{eqnarray}
\end{widetext}

The coefficients $f_i^j$ read:
\begin{widetext}
\begin{eqnarray}
f_\sigma^\sigma&=&\frac{1}{S} \, , 
f_b^\sigma=-f_\sigma^\sigma\left(\frac{\chi_1\chi_v^2}{B_\rho D_v}+\frac{\chi_1}{B_\rho}+\frac{\chi_2\chi_v}{D_v} \right) \, ,
f_\omega^\sigma=-f_\sigma^\sigma\left(\frac{\chi_2\chi_v^2}{B_vD_v}+\frac{\chi_2}{B_v}+\frac{\chi_1\chi_v}{D_v} \right) \\
f_\sigma^\omega&=&f_\sigma^\sigma\frac{B_\rho}{D_v}\left( \frac{\chi_1\chi_v}{B_\rho}+\chi_2\right) \, ,
f_b^\omega=S f_\sigma^\omega f_b^\sigma + \frac{\chi_v}{D_v} \, ,
f_\omega^\omega=S f_\sigma^\omega f_\omega^\sigma + \frac{B_\rho}{D_v} \\
f_\sigma^b&=&f_\sigma^\sigma\frac{B_v}{D_v}\left( \frac{\chi_2\chi_v}{B_v}+\chi_1\right) \, ,
f_b^b=S f_\sigma^b f_b^\sigma + \frac{B_v}{D_v} \, ,
f_\omega^b=S f_\sigma^b f_\omega^\sigma + \frac{\chi_v}{D_v}
\end{eqnarray}
\end{widetext}

with $D_v=B_vB_\rho-\chi_v^2$,  $
S=B_\sigma+\frac{\chi_1^2}{B_\rho}+\frac{\chi_2^2}{B_v}+\frac{\chi_1^2\chi_v^2}{D_v B_\rho}+\frac{\chi_2^2\chi_v^2}{D_v B_v}+2\frac{\chi_1\chi_2\chi_v}{D_v}
$,
and
$B_\sigma=\omega^2-k^2-m_{s,eff}^2$.

From Eq.~(\ref{matriz}), we get the following dispersion relation:
\begin{eqnarray} \label{dispersion}
(1&-&C_A^{ee}L_e)[(1+F^{pp}L_p+F^{nn}L_n \nonumber\\
&+&(F^{pp}F^{nn}-F^{pn}F^{np})L_pL_n]\nonumber\\
&-&C_A^{ep}C_A^{pe}L_pL_e(1+F^{nn}L_n)=0
\end{eqnarray}

The density fluctuations are given by
$$\delta\rho_i=\frac{3}{2}\frac{k}{P_{Fi}}\rho_{0i}A_{\omega i}.$$

At subsaturation densities, there are unstable modes identified by the sign of the imaginary
frequencies. For these modes, the growth rate is given by $\Gamma=-i\omega$.
The dynamical spinodal surface is defined by the region in ($\rho_p,\rho_n$) space, for a given wave vector $\mathbf k$ and
temperature $T$,  limited by the surface $\omega=0$.  In the  $k=0$ MeV limit, the thermodynamic
spinodal is obtained, defined by the surface in the ($\rho_p,\, \rho_n,\,
T$) space for which the curvature matrix of the free energy density is
zero, i.e., it has a zero eigenvalue. This relation has been discussed in more detail in Ref. \cite{ProvidenciaC-06}. 

\section{Results} \label{III}

In the present section, we discuss the effect of the two mixing terms $\omega\rho$ and $\sigma\rho$ on the crust-core transition properties and the importance of using an unified inner crust-core  EoS in the determination of the star radius. Starting  from the RMF models  NL3 \cite{Lalazissis-97}, TM1 \cite{Sumiyoshi-95}, and Z271 \cite{Horowitz01}, we build six families of models, each one having the same isoscalar properties, but varying the isovector properties through the mixed non-linear terms $\omega\rho$ and $\sigma\rho$, subsection \ref{families}. The properties of the models will be compared with present terrestrial and observational constraints and for each family the model(s) that satisfy these constraints will be identified. Next, we calculate the crust-core transition density and pressure for all the models,  applying  the Vlasov formalism, subsection \ref{vlasov}. Taking the two NL3 families, we  discuss the matching of the crust to the core to get the stellar matter EoS in subsection \ref{nl3}. For the inner crust, several possibilities will be considered.  Finally, we will apply the conclusions of subsection \ref{nl3}  to  construct the stellar matter EoS for the TM1 and Z271 families in section \ref{tm1-z271}, and we propose a set of procedures to build the stellar matter EoS from the knowledge of the crust-core density transition and the core EoS.

\subsection{Equation of state}
\label{families}

From all the terms presented in Eq. (\ref{eq:lnon-lin}), we will restrict our discussion to a set of models that have only one of the following  two non-linear coupling terms:
\begin{eqnarray} \label{wr-sr}
{\cal L}_{\omega \rho}&=&\Lambda_{\omega} g_v^2 g_\rho^2 \, V_{\mu}V^{\mu} \,\mathbf
b_{\mu}\cdot \mathbf b^{\mu}\\
{\cal L}_{\sigma \rho}&=&\Lambda_{\sigma} g_\sigma^2 g_\rho^2 \phi^2 \,\mathbf
b_{\mu}\cdot \mathbf b^{\mu}
\end{eqnarray}
to allow the modification of the density-dependence of the symmetry energy, by changing the $\rho-$meson effective mass \cite{Carriere03}, and discuss the star properties for different density dependences of the symmetry energy. 
The models NL3$\omega\rho$, Z271$\omega\rho$ and Z271$\sigma\rho$ were taken from \cite{Carriere03}. The others, namely, NL3$\sigma\rho$, TM1$\omega\rho$, and TM1$\sigma\rho$, are obtained by varying either the $\Lambda_{\omega}$ or $\Lambda_\sigma$ coupling constants, and by calculating the new $g_\rho$ constants, so that the symmetry energy at $\rho= 0.1$ fm$^{-3}$ has the same value as the reference model of the family.
 
Tables \ref{tab1} and \ref{tab2} show the  nuclear matter properties at the
saturation density for the models considered: the symmetry energy ${\mathcal{E}}_{{sym}}$,
the symmetry energy slope $L$, the symmetry energy curvature $K_{{sym}}$, and
$K_{\tau}=K_{sym}-6L-\frac{Q_0}{K}L$ (see Ref.\cite{Vidana09}). The tables also show the crust-core transition density, $\rho_{t}$, and pressure, $P_{t}$, calculated from the Vlasov method, as will be shown in the next subsection, and the neutron matter pressure, $P_N$, calculated at the nuclear saturation density.  Finally, the total binding energy per particle, the charge radii, the neutron radii, and the neutron skin thickness, $\Delta r_{\rm np}$, for the $^{208}$Pb nucleus are also given. We observe that the non-linear $\omega\rho$ and $\sigma\rho$ terms do not change the binding energy of the nuclei or the charge radius in more than 1\%.  Our results agree well with the ones in Refs. \cite{Carriere03, Horowitz01}. The  neutron radius, and therefore, also the neutron skin thickness decrease with decreasing L, for both families.

\begin{figure}
  \begin{tabular}{c}
\includegraphics[width=1\linewidth]{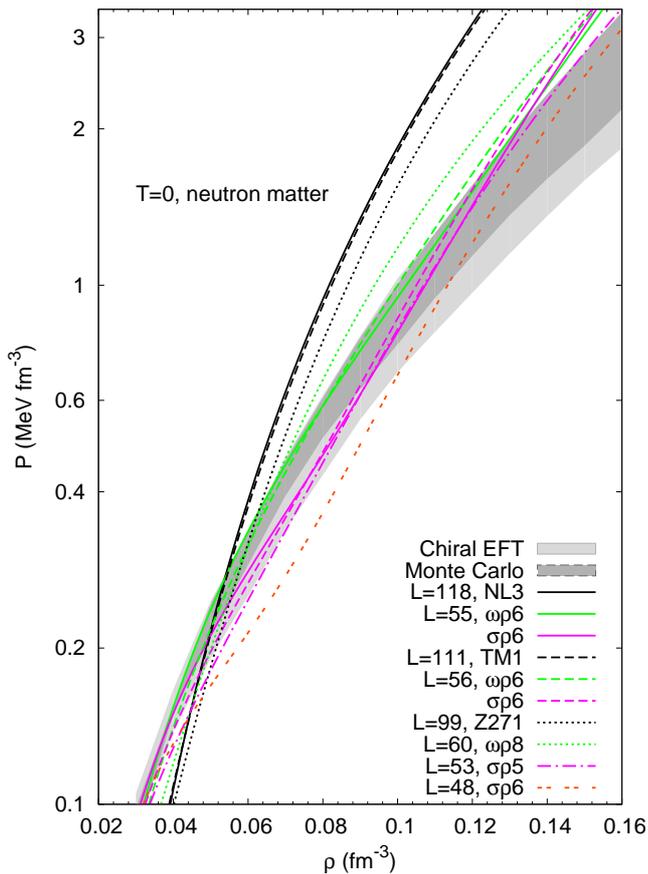}
  \end{tabular}
\caption{Neutron matter pressure, $P_N$, as a function of the density, for some of the models considered in this study. The coloured bands are the results from \cite{Hebeler-13} (light grey) and \cite{Gandolfi12} (dark grey).}
\label{fig1}
\end{figure}

In Fig. \ref{fig1}, we show the EoS of pure neutron matter for a set of models from tables \ref{tab1} and \ref{tab2}, and compare them with the result of microscopic calculations  based on nuclear interactions derived from chiral effective field theory (EFT) \cite{Hebeler-13}, or on realistic two- and three-nucleon interactions, using quantum Monte Carlo (MC) techniques \cite{Gandolfi12}. The band width indicates for each density the uncertainties of the calculation coming from the 3N interaction. At saturation density, this uncertainty is of the order of 5 MeV, $\sim 25-30\%$.

\begin{figure}
  \begin{tabular}{c}
\includegraphics[width=1\linewidth]{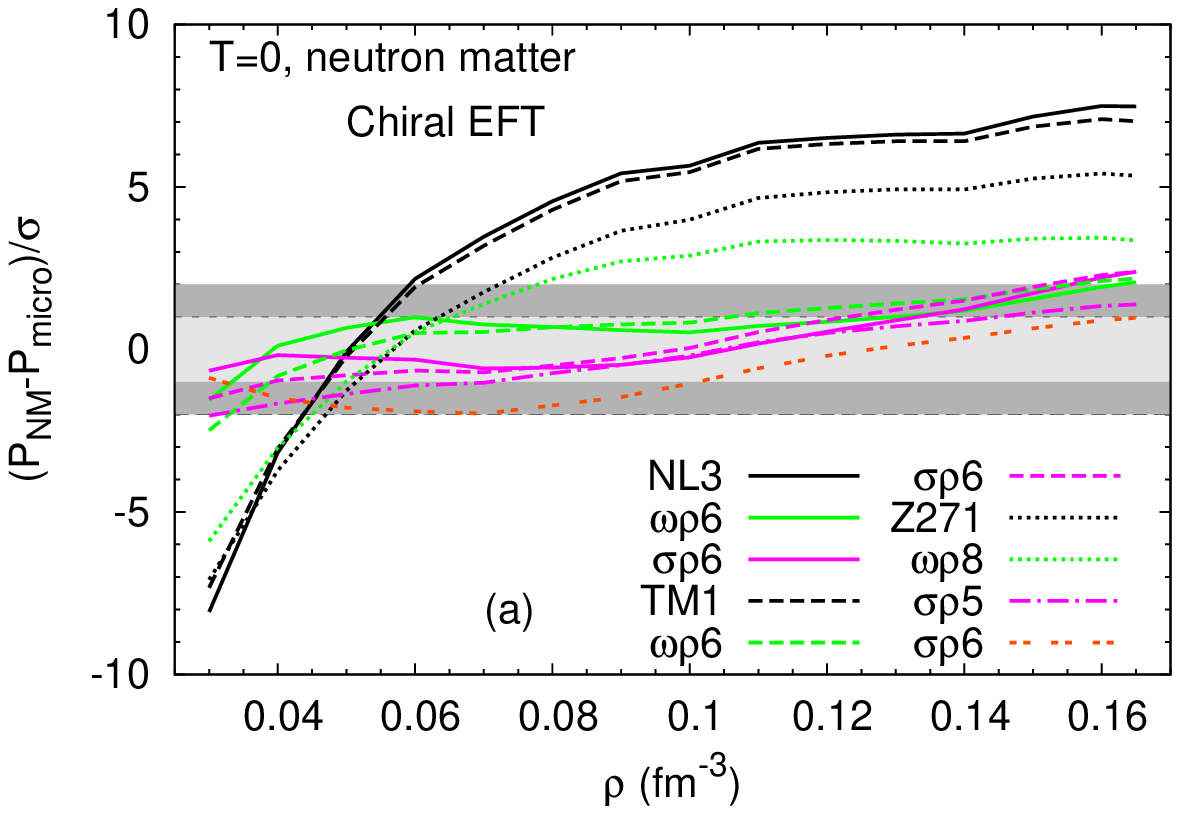}\\
\includegraphics[width=1\linewidth]{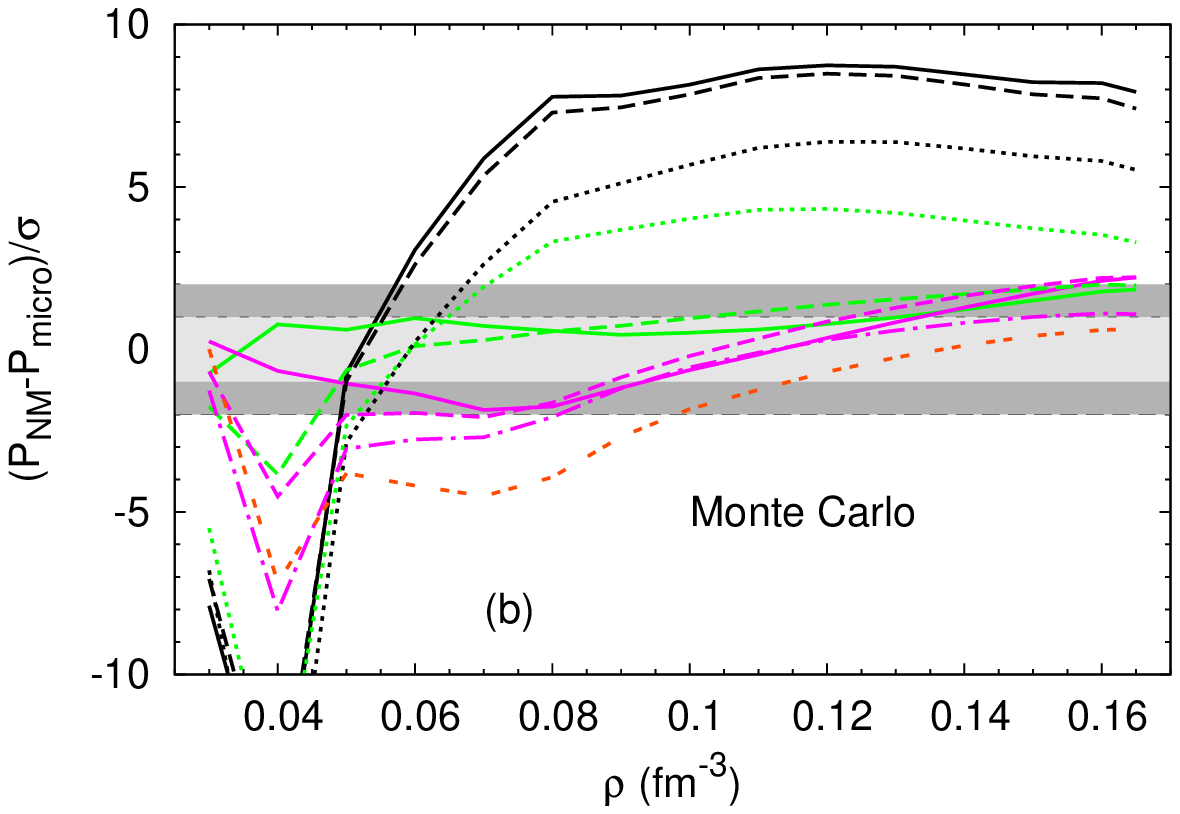}\\
  \end{tabular}
\caption{Difference between the neutron matter pressure for each model and the average pressure  obtained from a chiral EFT \cite{Hebeler-13} (a) and Monte Carlo \cite{Gandolfi12} (b) calculations, in units of the pressure uncertainty at each density, designated by $\sigma=\Delta P$. The grey bands represent  the calculation uncertainty (light) and twice this uncertainty (dark). See text for more details.}
\label{fig2}
\end{figure}

In  Fig. \ref{fig2} we plot the deviation of the neutron matter pressure for each model from the microscopic results of  Hebeler {\it et al.} \cite{Hebeler-13} and Gandolfi  {\it et al.}  \cite{Gandolfi12} data, in units of  the pressure uncertainty  $\Delta P$ of the microscopic calculations at each density, which we designate by $\sigma=\Delta P$.  The light grey bands represent the microscopic calculation uncertainties, meaning that the points inside those bands are within the data limits. The dark grey bands correspond to twice the calculation uncertainties, $2\sigma$. Except for the head of each family and Z271$\omega\rho$8,   all other models  agree well  with the results of  \cite{Hebeler-13}, and above $\rho=0.05$ fm$^{-3}$ with the MC results  \cite{Gandolfi12}.

Models TM1, NL3 and Z271 have a too stiff neutron matter EoS. These models do not satisfy several other constraints imposed by experiments, see \cite{Dutra-14}, in particular, the symmetry energy and/or its slope is too high, as well as the incompressibility. However, including the $\omega\rho$ or the $\sigma\rho$ terms, makes  the symmetry energy  softer, and  we obtain some models that satisfy most of the constraints imposed in \cite{Dutra-14}. We will not consider the  $K_{\tau}$ constraint included in \cite{Dutra-14} because it has a large uncertainty associated to it. 
 The NL3x and Z271x models have an incompressibility at saturation, $K$,  just above the upper limit used in \cite{khan12,Dutra-14}, $K=230\pm 40$ MeV,  and TM1 only satisfies this constraint  within 10\%.  However, the incompressibility of these models is within the range $250<K<315$ MeV predicted in \cite{Stone14,Stone16}.  Therefore, we consider that the $K$ constraint is satisfied.

\begin{figure}
  \begin{tabular}{c}
\includegraphics[width=1\linewidth]{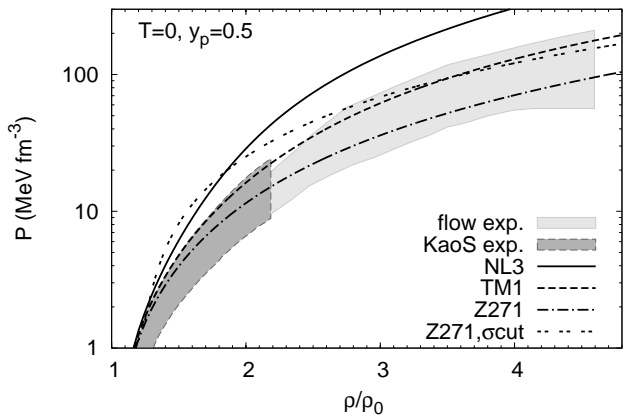}
  \end{tabular}
\caption{Symmmetric matter pressure, $P$, as a function of the density, for the models considered in this study. The coloured bands are the experimental results obtained from collective flow data in
heavy-ion collisions \cite{Danielewicz-02}  (light grey) and from the KaoS experiment \cite{Lynch09,Fuchs06} (dark grey).}
\label{fig3}
\end{figure}

Constraints at suprasaturation densities from heavy ion collisions were also considered in \cite{Dutra-14}. For reference,  in Fig. \ref{fig3}, we show the symmetric matter pressure as a function of the density for the models considered in this study. This plot also  includes a modified Z271 model, which  has an extra effective potential that will be discussed later.
 We compare the different EoS with  the experimental results obtained from collective flow data in heavy-ion collisions \cite{Danielewicz-02} and from the KaoS experiment \cite{Lynch09,Fuchs06}. Only Z271 and TM1 satisfy the constraints.

Summarizing the above discussion, models TM1$\omega\rho6$, TM1$\sigma\rho6$, and Z271$\sigma\rho5-6$ satisfy the constraints imposed in Ref. \cite{Dutra-14}, except the one in $K_\tau$, the constraints from neutron matter calculations and the ones from Refs. \cite{Stone16, Stone14}. Models NL3$\omega\rho$6 and NL3$\sigma\rho6$ only fail the flow and KaoS experiments.

\subsection{Crust-core transition}
\label{vlasov}

\begin{figure}[t]
  \begin{tabular}{c}
\includegraphics[width=1\linewidth]{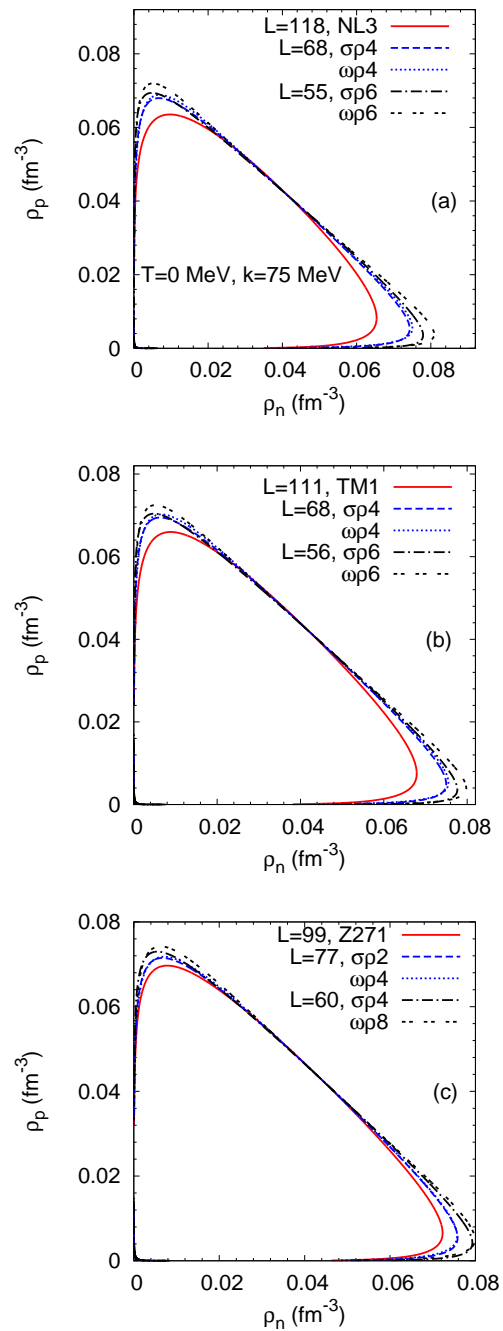}
  \end{tabular}
\caption{Spinodal regions for the models considered.}
\label{fig4}
\end{figure}

In Fig. \ref{fig4}, the dynamical spinodals for the six families under study are represented. They have been obtained solving the dispersion relation (\ref{dispersion}) for a zero energy mode, $\omega=0$, and taking the wave number $k=75$ MeV. We have considered this value of $k$ because the extension of the spinodal section is close to the envelope of all spinodal sections. For NL3 and TM1, we have considered, besides the head of the family, the two parametrizations with  $\omega\rho$ or $\sigma\rho$ terms that give $L=55$ and 68 MeV; for Z271, we take the models with $L=76$ and 60 MeV. All the values of $L$ chosen are within the different constraints imposed in \cite{Dutra-14} for $L$. Some conclusions may be drawn from the figure: a) the larger $L$, the smaller the spinodal section, as discussed in previous works with different models \cite{Pais10}; b) the term $\omega\rho$ makes the spinodal section larger, except for the very large isospin asymmetries, both very neutron rich or very proton rich.

\begin{figure}
  \begin{tabular}{c}
\includegraphics[width=1\linewidth]{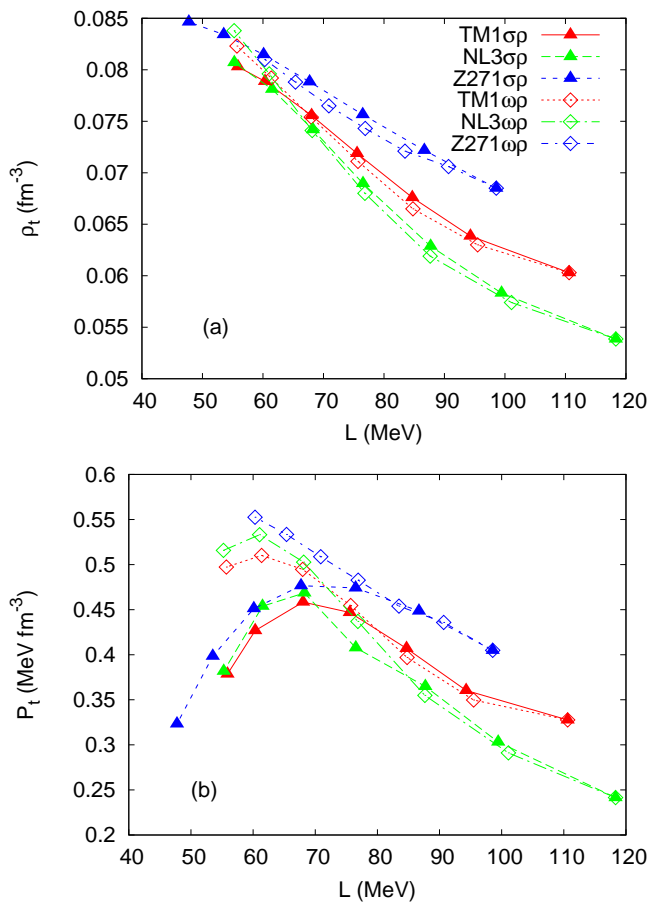}
  \end{tabular}
\caption{Crust-core transition (a) densities, $\rho_t$, and (b) pressure, $P_t$, as a function of the slope of the symmetry energy, $L$. }
\label{fig5}
\end{figure}
 
 The crust-core  transition density, $\rho_t$, is calculated from the crossing between the spinodal sections and the EoS for $\beta$-equilibrium matter. These densities are given in Tables \ref{tab1} and \ref{tab2} for all models under study, and are represented as a function of $L$ in Fig. \ref{fig5}, top panel.  The crossing between the $\omega\rho$ and $\sigma\rho$  spinodals,  for a given $L$, occurs close to the crossing of the $\beta$-equilibrium EoS with the spinodals and, therefore, the transition densities do not differ much, whether taking the $\omega\rho$ or the $\sigma\rho$ term. A difference, however, is seen when the transition pressures, $P_t$, are compared, see Fig. \ref{fig5}, bottom panel: while for the $\omega\rho$ terms, the transition pressure decreases monotonically with $L$, the exception being the lowest $L$ values for the TM1 and NL3 families, the $\sigma\rho$ terms gives rise to an increase of $P_t$ with $L$, until $L=70-80$ MeV, and only above this value, the pressure decreases with an increase of $L$.

\begin{figure}
  \begin{tabular}{c}
\includegraphics[width=1\linewidth]{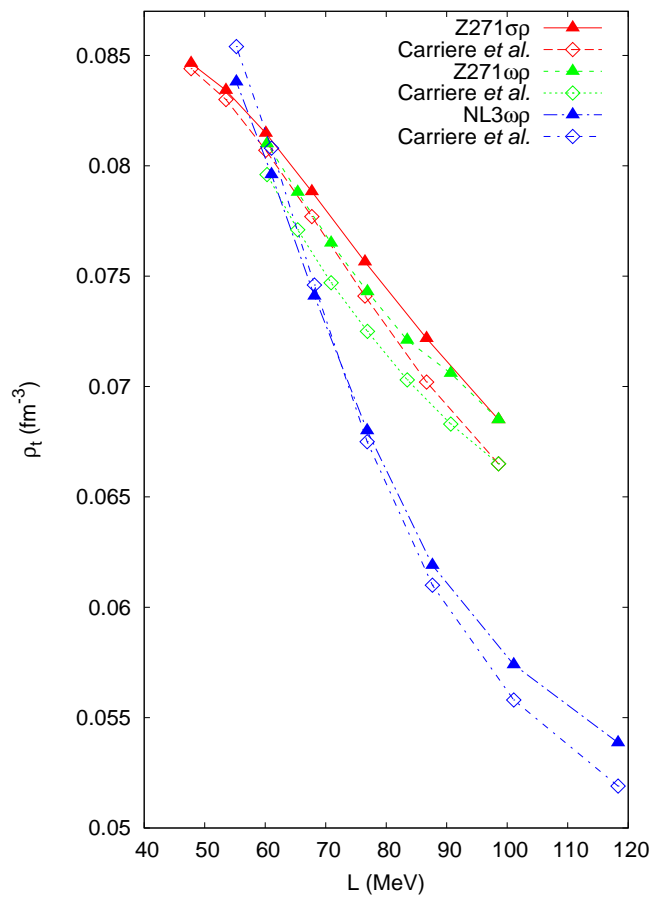}
  \end{tabular}
\caption{Crust-core transition densities, $\rho_t$, as a function of the slope of the symmetry energy, $L$, where we compare with the results obtained in Ref. \cite{Carriere03}. }
\label{fig6}
\end{figure}

For values of $L$ below 80 MeV, $P_t$ is generally larger in the models with the $\omega\rho$ term, and this difference may be as large as 25\% for the smaller $L$ shown. This difference has direct implications in the moment of inertia of the crust, which is proportional to $P_t$, $I_{crust} \sim \frac{16\pi}{3}\frac{R_t^6P_t}{R_s}$ in lowest order  \cite{Link-99,Shen-10,Piekarewicz-14},  where $R_t$ is the crust thickness, and $R_s$ is the Schwarzschild radius. Besides the transition pressure, the moment of inertia of the crust also depends on the crust thickness. In the following,  we will also see that the $\sigma\rho$ terms give rise to smaller  crust thicknesses.  This implies that smaller crust moments of inertia are expected, if the $\sigma\rho$ term is used to modify the symmetry energy.

In Fig. \ref{fig6}, we compare our results for the crust-core transition density with the ones obtained in Ref. \cite{Carriere03}, calculated within a relativistic random-phase-approximation \cite{horowitz91}, for the Z271$\sigma\rho$ and $\omega\rho$ models, and for the NL3$\omega\rho$ family.  For the Z271* models, we obtain slightly larger values for $\rho_t$, and this difference increases with increasing $L$, but even for the largest value of $L$, the difference is below  5\%. For the NL3$\omega\rho$ models, the behaviour is slightly different: for $L<70$ MeV, our results are below the ones of Ref. \cite{Carriere03}, and for $L>70$ MeV, we get larger values, though the overall difference between them is below 5\%.

\subsection{Mass-radius curves for the NL3 families}
\label{nl3}

\begin{figure}
   \begin{tabular}{c }
\includegraphics[width=1\linewidth]{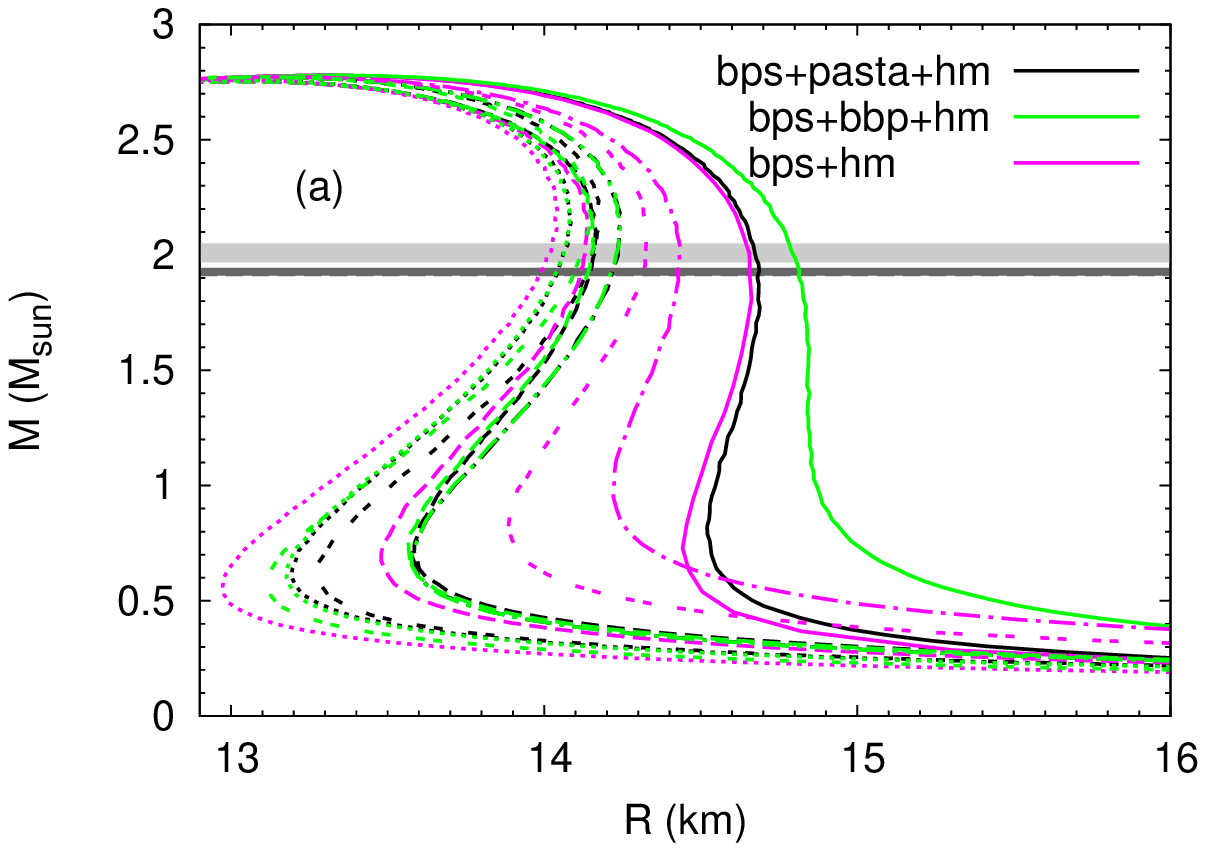}\\
 \includegraphics[width=1\linewidth]{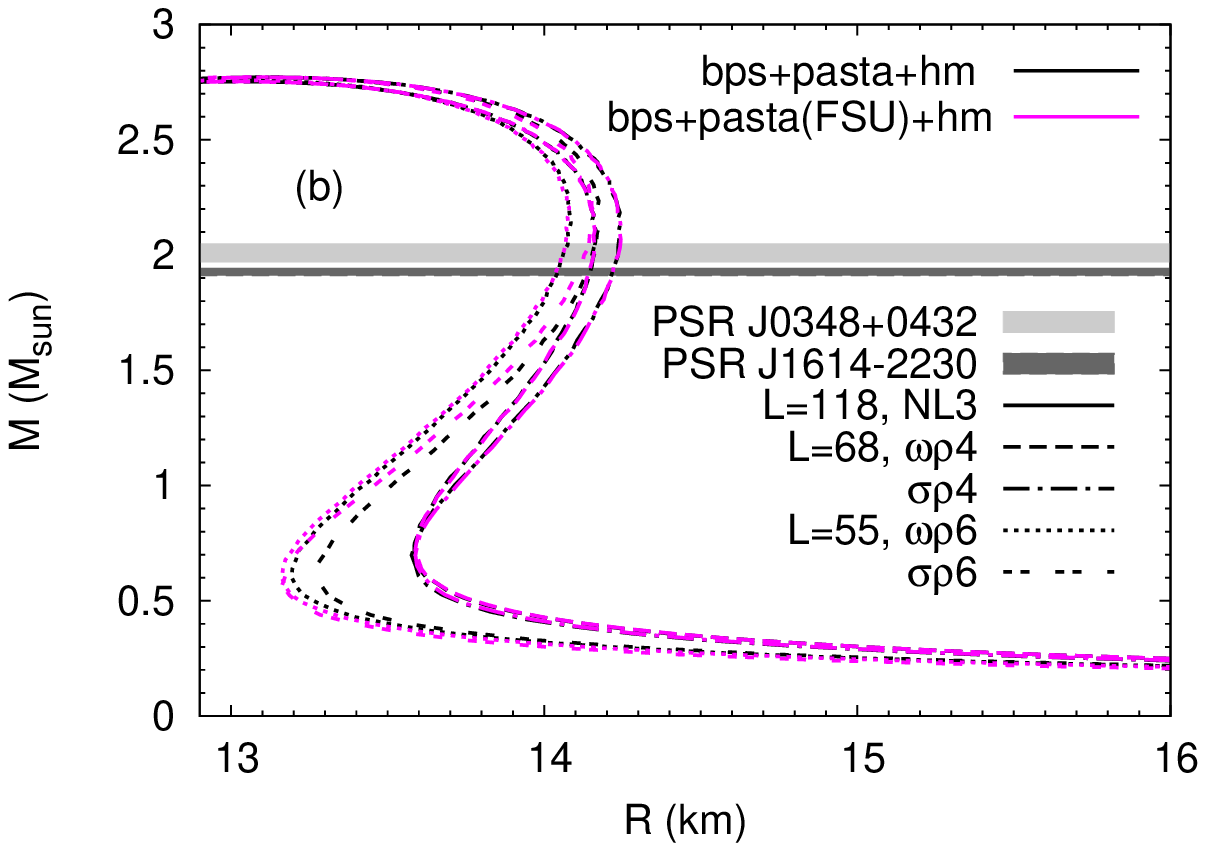}\\
   \end{tabular}
\caption{$M(R)$ relations for the NL3 models: (a) BPS EoS \cite{BPS-71} for the outer crust and the homogeneous matter EoS for the inner crust and core, bps+hm, (pink lines); the BPS+BBP \cite{Glendenning00} for the outer crust, and the homogeneous matter EoS for the inner crust and core, bps+bbp+hm,  (green lines); and the BPS for the outer crust, the pasta configurations calculated from TF for the inner crust and the homogeneous matter EoS for the core, bps+pasta+hm (black lines); (b) BPS EoS for the outer crust, the pasta calculated from TF  for the respective models (black lines), or  for the FSU model (pink lines), for the inner crust, and the homogeneous matter EoS for the core. The horizontal bands indicate the mass uncertainties associated to the PSR J0348+0432 \cite{Antoniadis13} and PSR J1614-2230 \cite{Fonseca16}  masses.}   
\label{fig7}
\end{figure}

We calculate the mass-radius curves for the EoS under study, by  integrating the Tolmann-Oppenheimer-Volkof equations \cite{tolman39,oppenheimer39}, for relativistic spherical stars in hydrostatic equilibrium.  In Fig. \ref{fig7}, the curves for the NL3 family are shown. In order to construct the EoS of stellar matter, we take,  besides the EoS of the core, the  BPS EoS for the outer crust, and several models for the inner crust: a) for models NL3$\omega\rho$ and $\sigma\rho$, we calculate the inner crust, within a Thomas-Fermi calculation \cite{Avancini-08,Grill-14}. These inner crust EoS will also be considered when building the complete stellar matter EoS, within other models with similar properties; b) as an alternative that tests the use of a non-unified EoS, we consider the FSU inner crust EoS \cite{Grill-14}, between the neutron drip density and the crust-core transition density, calculated within the dynamical spinodal method, for models with a similar slope $L$; c) we match directly the core EoS to the outer crust BPS EoS; d) the BPS plus Bethe-Baym-Pethick (BBP) EoS for densities below 0.01 fm$^{-3}$ is matched directly to the core EoS,  as suggested in \cite{Glendenning00}. Finally, we will estimate the error  we introduce in quantities, such as the radius and mass, if the unified inner crust EoS is not used.

\begin{figure}
  \begin{tabular}{c}
\includegraphics[width=1\linewidth]{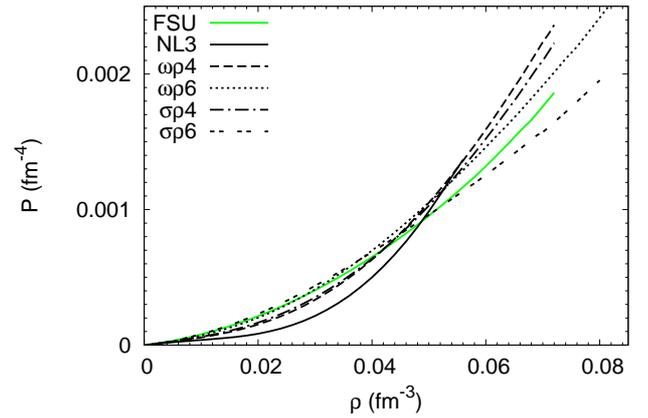}
  \end{tabular}
\caption{ Inner crust EoSs considered in the present study. }
\label{fig8}
\end{figure}

In Fig. \ref{fig8} and in the Appendix, we show the inner crust EoSs that we are using in this paper.  Both the FSU,  the NL3, and the NL3$\omega\rho6$ inner crusts are  given in \cite{Grill-14}.
In the top panel of Fig. \ref{fig7},  we compare the mass-radius curves for the stellar EoS obtained using the scenarios a) BPS plus an unified inner crust and core EoS (bps+pasta+hm), b) the inner crust EoS is replaced by the homogeneous matter EoS  (bps+hm) and c) the BBP EoS is used for the low density  inner crust EoS, and a transition to homogeneous matter occurs at $\sim 0.01$ fm$^{-3}$, below the crust-core transition (bps+bbp+hm). Totally neglecting the inner crust EoS (dashed curves) is a quite rough approximation for all EoS, except for NL3. Although the effect on the maximum mass is negligible, the same is not true for the radius. The families of stars with an unified inner crust-core EoS (solid lines) have larger (smaller) radii than the configurations without inner crust (dashed lines) for the NL3$\omega\rho$ (NL3$\sigma\rho$) models, see Tables \ref{tab3} and \ref{tab4}.  Including the BBP EoS between the neutron drip and $\rho=0.01$ fm$^{-3}$ (dotted lines) will generally reduce the differences with respect to the unified EoS, although the improvement depends a lot on the model. 

Fig. \ref{fig7} allows a comparison between mass-radius curves obtained with EoS whose density dependence of the symmetry energy is modified by means of a mixing $\omega\rho$ or $\sigma\rho$ term in the Lagrangian density.  Within models with the same $L$, the $\sigma\rho$ models give slightly larger radii, the differences being larger for $M \gtrsim 1.4 M_\odot$.  For a $ 1.4 M_\odot$ star, we have obtained a difference of $\sim 100$ m. These differences reflect themselves in the crust thickness, see Tables \ref{tab3} and \ref{tab4}. The most critical approximation, giving rise to the largest error, occurs when the inner crust is completely neglected.

\begin{figure}
   \begin{tabular}{c}
\includegraphics[width=1\linewidth]{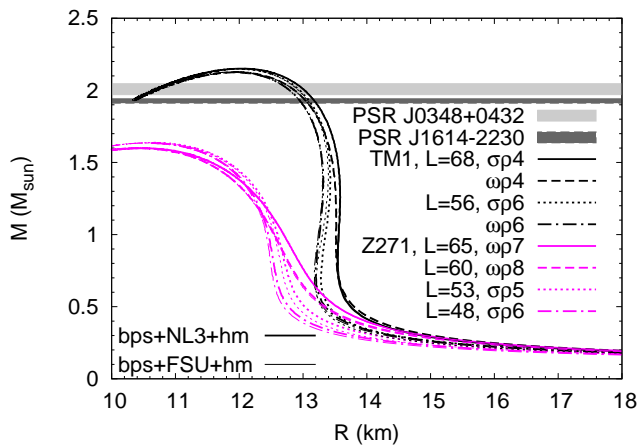}
   \end{tabular}
\caption{$M(R)$ relations for the all the TM1 and Z271 models considered: the stellar matter EoS contains  the BPS outer crust EoS, the FSU inner crust  (thin lines), the NL3$\omega\rho$ or the $\sigma\rho$ (thick lines) pasta phase EoS, and the core homogeneous matter EoS. }
\label{fig9}
\end{figure}

\begin{figure}
  \begin{tabular}{c}
\includegraphics[width=1\linewidth]{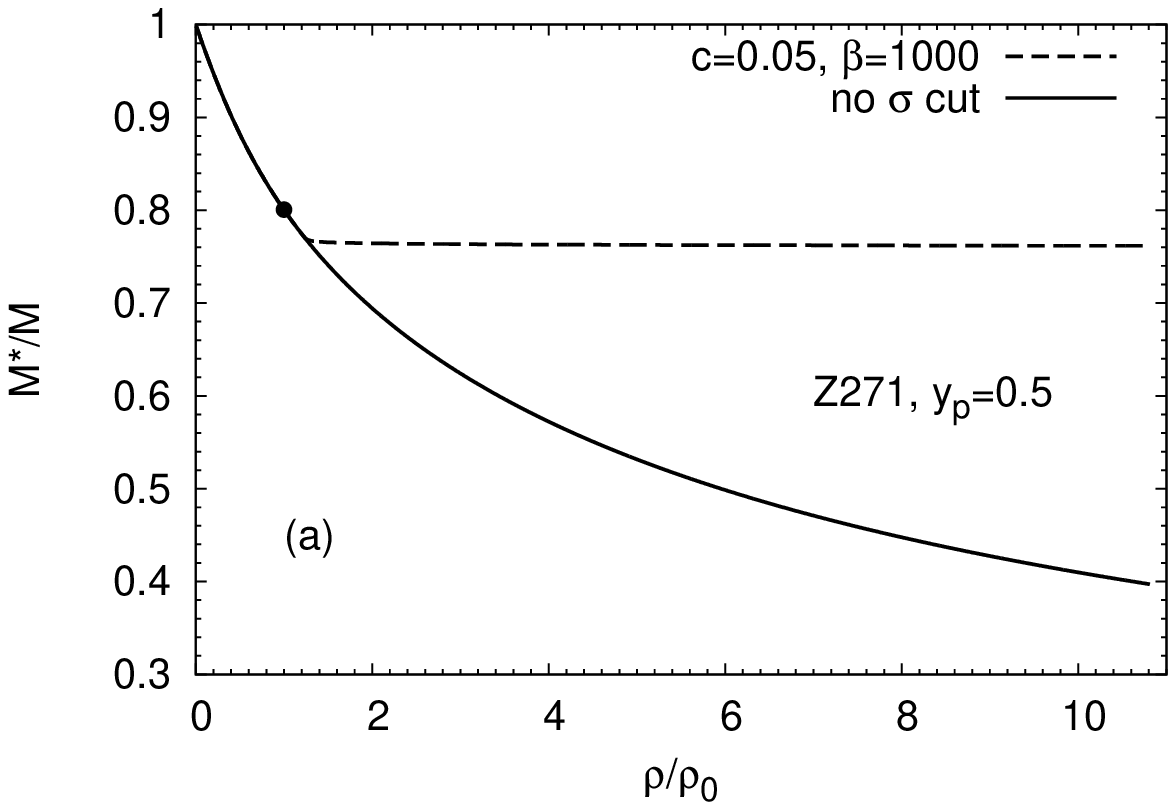}\\
\includegraphics[width=1\linewidth]{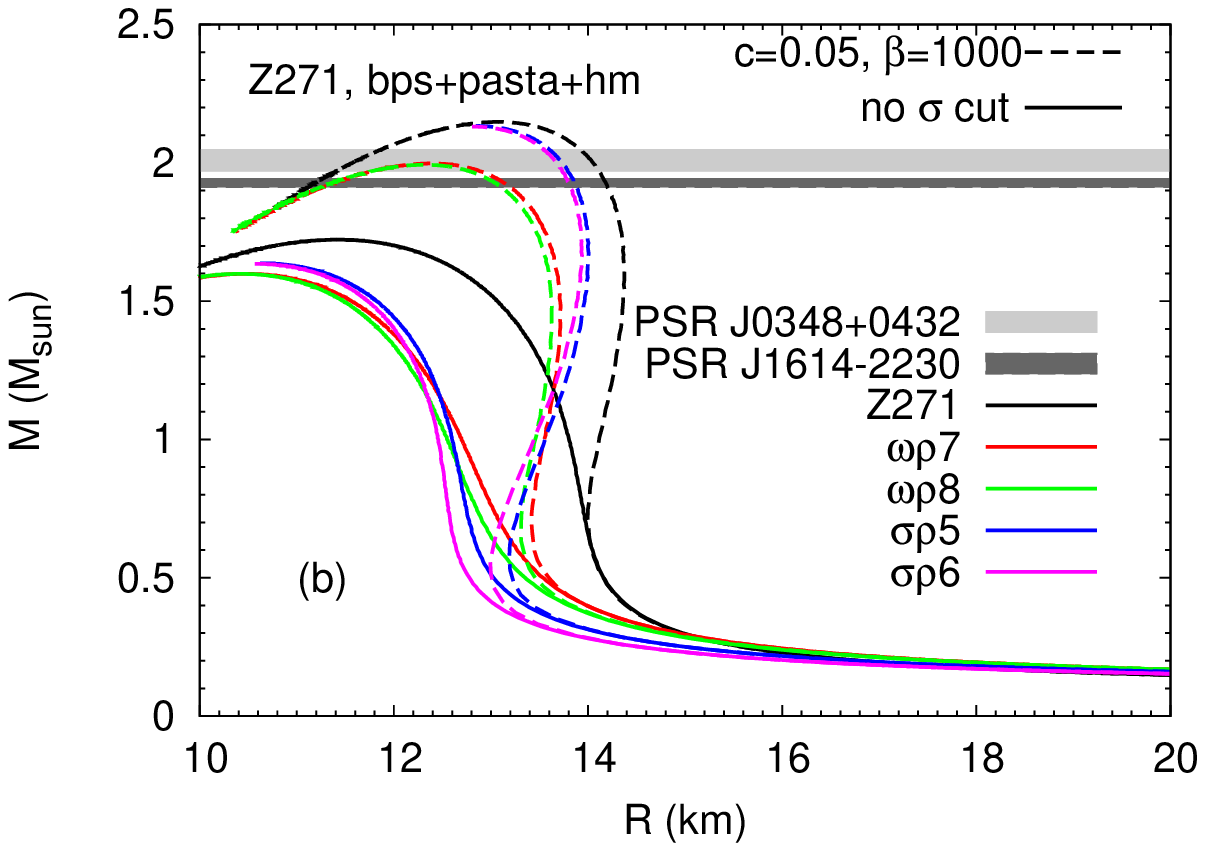}
  \end{tabular}
\caption{ (a) Nucleon effective mass as a function of the density for isospin symmetric matter for the Z271 model with (dashed line) and without (solid line) the $\sigma$ cut potential. The black dot shows the effective mass at the nuclear saturation density, $\rho_0$; (b) $M(R)$ relations for the Z271 set of models, with (dashed lines) and without (solid lines) the $\sigma$ cut potential. }
\label{fig10}
\end{figure}

\subsection{Stellar matter EoS}
\label{tm1-z271}

Most of the times, the inner crust EoS calculated within the same model is not available. In the previous section, we have tested  the implications of not including the inner crust EoS, or only part of it.  We now discuss the possible use of a inner crust EoS obtained for a different model. 
 In the bottom panel of Fig. \ref{fig7}, we plot the mass-radius curves of stars obtained, considering for the inner crust the unified pasta calculation as before, and the inner crust EoS calculated for FSU, a model with the symmetry energy slope $L=60$ MeV, not far from the slope $L$ of  NL3$\omega\rho4,6$ and  NL3$\sigma\rho4,6$. Except for  NL3$\sigma\rho6$, where we obtain a difference of $\sim 50$ m ($\sim 40$ m) for a 1$M_\odot$  ( 1.4$M_\odot$) star, and for NL3$\omega\rho6$, where we obtain $\sim 20$ m for a 1$M_\odot$ star, the error on the determination of the radius is negligible for all masses.  This can be better seen in Tables \ref{tab3} and \ref{tab4}.

\begin{table*}
\caption{The radius and crust thickness of a  1.4 $M_\odot$ star, computed with an inner crust calculated within NL3$\sigma\rho$/NL3$\omega\rho$ and FSU, or using the BBP EoS, or  without an inner crust (\textit{no ic}) are shown. In all cases, we consider the BPS EoS for the outer crust. }  \label{tab3}
  \begin{tabular}{c c c c c c c c c c c c c c c}
    \hline
    \hline
 Model & \phantom{a} & $L$ (MeV) & \phantom{a} & $R_{1.4 M\odot}$ & $\Delta R_{crust}$ (NL3) & \phantom{a} & $R_{1.4M\odot}$ &$\Delta R_{crust}$ (FSU) & \phantom{a} & $R_{1.4M\odot}$ & $ \Delta R_{crust}$ (BBP) & \phantom{a} & $R_{1.4M\odot}$ & $\Delta R_{crust}$ (no ic) \\
    \hline

NL3 &\phantom{a}& 118 &\phantom{a}& 14.630 & 1.325 &\phantom{a} & - & - & \phantom{a} & 14.847 & 1.508 & \phantom{a} & 14.594 & 1.249 \\

$\omega\rho$4 &\phantom{a}& 68 &\phantom{a}& 13.928 & 1.450 &\phantom{a} & 13.928 & 1.377 & \phantom{a} & 13.916 & 1.453 & \phantom{a} & 13.879 & 1.416 \\

$\omega\rho$6 &\phantom{a}& 55 &\phantom{a}& 13.753 & 1.425 & \phantom{a} & 13.745 & 1.432 & \phantom{a} & 13.749 & 1.432 & \phantom{a} & 13.671 & 1.355 \\

$\sigma\rho$4 &\phantom{a}& 68 &\phantom{a}& 13.982 & 1.441 & \phantom{a} & 13.983 & 1.388 &\phantom{a}& 13.977 & 1.445 & \phantom{a} & 14.313 & 1.781 \\

$\sigma\rho$6 &\phantom{a}& 55 &\phantom{a}& 13.846 & 1.400 & \phantom{a} & 13.806 & 1.408 & \phantom{a} & 13.785 & 1.334 &\phantom{a}& 14.117 & 1.666 \\

    \hline
    \hline
  \end{tabular}
\end{table*}

\begin{table*}
\caption{The radius and crust thickness of a  1.0 $M_\odot$ star, computed with an inner crust calculated within NL3$\sigma\rho$/NL3$\omega\rho$ and FSU, or using the BBP EoS, or without an inner crust (\textit{no ic}), are shown. In all cases, we consider the BPS EoS for the outer crust. }  \label{tab4}
  \begin{tabular}{c c c c c c c c c c c c c c c}
    \hline
    \hline
 Model & \phantom{a} & $L$ (MeV) & \phantom{a} & $R_{1 M\odot}$ & $\Delta R_{crust}$ (NL3) & \phantom{a} & $R_{1 M\odot}$ &$\Delta R_{crust}$ (FSU) & \phantom{a} & $R_{1 M\odot}$ & $ \Delta R_{crust}$ (BBP) & \phantom{a} & $R_{1 M\odot}$ & $\Delta R_{crust}$ (no ic) \\
    \hline

NL3 &\phantom{a}& 118 &\phantom{a}& 14.547 & 1.956 &\phantom{a} & - & - & \phantom{a} & 14.870 & 2.230 & \phantom{a} & 14.489 & 1.850 \\

$\omega\rho$4 &\phantom{a}& 68 &\phantom{a}& 13.681 & 2.077 &\phantom{a} & 13.686 & 1.979 & \phantom{a} & 13.665 & 2.079 & \phantom{a} & 13.621 & 2.035\\

$\omega\rho$6 &\phantom{a}& 55 &\phantom{a}& 13.423 & 2.020 & \phantom{a} & 13.402 & 2.020 & \phantom{a} & 13.410 & 2.025 & \phantom{a} & 13.300 & 1.913 \\

$\sigma\rho$4 &\phantom{a}& 68 &\phantom{a}& 13.713 & 2.057& \phantom{a} & 13.720 & 1.99 &\phantom{a}& 13.710 & 2.067 & \phantom{a} & 14.223 & 2.581 \\

$\sigma\rho$6 &\phantom{a}& 55 &\phantom{a}& 13.511 & 1.987 & \phantom{a} & 13.457 & 1.993 & \phantom{a} & 13.423 & 1.885 &\phantom{a}& 13.924 & 2.386\\

    \hline
    \hline
  \end{tabular}
\end{table*}

\begin{table*}[htb]
\caption{ Maximum mass properties.   For the TM1* and Z271* models, we used for the inner crust FSU, and NL3$\omega\rho$ and $\sigma\rho$. For TM1 and Z271, we used TM1. The radii, $R$, are given in km.}  \label{tab5}
  \begin{tabular}{c c c c c c c cccccccccccccccccc}
    \hline
    \hline
 Model& \phantom{a} & $L$ (MeV) & \phantom{a} & $M_g$ (M$_{\odot}$) & \phantom{a} & $M_b$ (M$_{\odot}$) & \phantom{a} & $R$ (BPS) & \phantom{a} & $R$ (FSU) & \phantom{a} & $R$ (NL3) & \phantom{a} & $\epsilon_0$ (fm$^{-4}$) & \phantom{a} & $\rho_c$ (fm$^{-3}$) \\
    \hline

NL3 & \phantom{a} & 118 & \phantom{a} & 2.779 & \phantom{a} & 3.384 & \phantom{a} & 13.289 & \phantom{a} & - & \phantom{a} & 13.300& \phantom{a} & 4.409 &\phantom{a} & 0.669  \\

$\omega\rho$4 &\phantom{a}& 68 &\phantom{a}& 2.753 &\phantom{a}& 3.376 & \phantom{a} & 13.017 &\phantom{a}& 13.031 & \phantom{a} & 13.022 &\phantom{a}& 4.520 &\phantom{a}& 0.689 \\

$\omega\rho$6 &\phantom{a}& 55 &\phantom{a}& 2.758 &\phantom{a}& 3.391 & \phantom{a} & 12.994 &\phantom{a}& 13.005& \phantom{a} & 13.014 &\phantom{a}& 4.496 &\phantom{a}& 0.687 \\

$\sigma\rho$4 &\phantom{a}& 68 &\phantom{a}& 2.771 &\phantom{a}& 3.400 & \phantom{a} & 13.190 &\phantom{a}& 13.116 & \phantom{a} & 13.114 &\phantom{a}& 4.450 &\phantom{a}& 0.679 \\

$\sigma\rho$6 &\phantom{a}& 55 &\phantom{a}& 2.773 &\phantom{a}& 3.409 & \phantom{a} & 13.141 &\phantom{a}& 13.070 & \phantom{a} & 13.079 &\phantom{a}& 4.469 &\phantom{a}& 0.682  \\

\hline
TM1 & \phantom{a} & 111 & \phantom{a} & 2.183 & \phantom{a} & 2.544 & \phantom{a} & 12.494 & \phantom{a} & 12.386 & \phantom{a} &- & \phantom{a} & 5.345 &\phantom{a} & 0.851 \\

$\omega\rho$4 &\phantom{a}& 68 &\phantom{a}& 2.125 &\phantom{a}& 2.488 & \phantom{a} & 12.140 &\phantom{a}& 11.950 &\phantom{a}& 11.955 & \phantom{a} &5.653 &\phantom{a}& 0.904 \\

$\omega\rho$6 &\phantom{a}& 56 &\phantom{a}& 2.127 &\phantom{a}& 2.496 & \phantom{a} & 11.973 &\phantom{a}& 11.892 &\phantom{a}& 11.893 & \phantom{a} &5.663 &\phantom{a}& 0.907 \\

$\sigma\rho$4 &\phantom{a}& 68 &\phantom{a}& 2.150 &\phantom{a}& 2.521 & \phantom{a} & 12.105 &\phantom{a}& 12.034 &\phantom{a}& 12.038 & \phantom{a} &5.565 &\phantom{a}& 0.889 \\

$\sigma\rho$6 &\phantom{a}& 56 &\phantom{a}& 2.148 &\phantom{a}& 2.523 & \phantom{a} & 12.019 &\phantom{a}& 11.977 &\phantom{a}& 11.992 & \phantom{a} &5.578 &\phantom{a}& 0.893  \\
\hline

Z271 & \phantom{a} & 99 & \phantom{a} & 1.722 & \phantom{a} & 1.944 & \phantom{a} & 11.554 & \phantom{a} & 11.425 & \phantom{a} & - & \phantom{a} &6.470 &\phantom{a} & 1.066  \\

$\omega\rho$7 &\phantom{a}& 65 &\phantom{a}& 1.599 &\phantom{a}& 1.807 & \phantom{a} & 10.544 &\phantom{a}& 10.466 &\phantom{a}& 10.466 & \phantom{a} & 7.804 &\phantom{a}& 1.277 \\

$\omega\rho$8 &\phantom{a}& 60 &\phantom{a}& 1.599 &\phantom{a}& 1.809 & \phantom{a} & 10.435 &\phantom{a}& 10.406 &\phantom{a}& 10.417 & \phantom{a} &7.850 &\phantom{a}& 1.285 \\

$\sigma\rho$5 &\phantom{a}& 53 &\phantom{a}& 1.637 &\phantom{a}& 1.857 & \phantom{a} & 10.733 &\phantom{a}& 10.611 &\phantom{a}& 10.621 & \phantom{a} &7.422 &\phantom{a}& 1.220 \\

$\sigma\rho$6 &\phantom{a}& 48 &\phantom{a}& 1.635 &\phantom{a}& 1.857 & \phantom{a} & 10.653 &\phantom{a}& 10.553 &\phantom{a}& 10.571 & \phantom{a} &7.463 &\phantom{a}& 1.228  \\

    \hline
    \hline
  \end{tabular}
\end{table*}

The above discussion indicates that the determination of the radii of stars requires that some care is taken when matching the crust EoS to the core EoS. Non-unified EoS may give rise to large uncertainties. It is possible, however, to build an adequate  non-unified EoS, if the inner crust EoS is properly chosen. We have shown that taking the inner crust EoS of a model with similar symmetry energy properties, as the ones of the EoS used for the core, allowed the determination of the radii of the family of stars with masses above $1\, M_\odot$ with an uncertainty below 50 m. The inclusion of the inner crust EoS has definitely a strong effect on the radius of low and intermediate mass neutron stars.

For the TM1 and Z271 families, we do not have an unified EoS for the inner crust. In order to test the above conclusion, we have built the stellar EoS taking for the inner crust, between the neutron drip density and the crust-core transition density, and calculated within the dynamical spinodal method, a) the FSU EoS (thin lines) and b) the EoS obtained for the NL3$\omega\rho$  (dashed lines) and NL3$\sigma\rho$ (solid lines) models, choosing the EoS that has the properties closer to the ones of each one of the models of the TM1 or Z271 familes (thick lines), see  Fig. \ref{fig9}. Table \ref{tab5} gives the correspondent maximum mass properties. In some cases, the curves (almost) coincide: this occurs for all models with $L> 60$ MeV. Considering the two inner crust EoS, small differences occur for models with $L< 60$ MeV, but the differences for the 1$M_\odot$ (1.4$M_\odot$) are never larger than 50 (30) m. This procedure to choose the inner crust EoS seems to be a quite robust alternative to the unified inner crust EoS.

Although the SNM and PNM properties of some of the  Z271 models are in good agreement with experimental results or microscopic calculations,  they predict a too small maximum mass star. This problem can be solved by introducing an extra effective potential, dependent on the $\sigma$ meson, that hinders the effective nucleon mass to stop decreasing at a density above the saturation density, as suggested in \cite{Maslov-15}. In  Figure \ref{fig10},  we show, in the top panel, the nucleon effective mass as a function of the density for the Z271 model, with and without the $\sigma$ cut potential, $\Delta U_f$, eq. 10 in Ref. \cite{Maslov-15}, written as
\begin{eqnarray}
\label{sigmacut}
\Delta U_f&=&\alpha \ln(1+\exp\left[\beta(g_s\phi-f)\right]), \nonumber \\
f&=&f_0+c_{\sigma}(1-f_0).
\end{eqnarray}

We have used the parameters $\beta=1000$ and $c_\sigma=0.05$ to be able to get maximum masses of, at least, 2 $M_\odot$, as shown in the bottom panel (see also Ref. \cite{Dutra16}, where they choose different parameters).  $f_0$ and $\alpha$ have the same values as in Ref.\cite{Maslov-15},  0.2 and 4.822$\times 10^{-4}$ $m_N^4$, respectively. The stellar matter EoS were built using the BPS outer crust EoS, the most adequate  NL3$\omega\rho$ or NL3$\sigma\rho$  inner crust EoS, and the homogeneous matter core EoS. The potential given in Eq. (\ref{sigmacut}) does not allow  constructing a EoS that simultaneously satisfies the $2M_\odot$ and the KaoS constrains, see Fig. \ref{fig3}.

\begin{figure}
  \begin{tabular}{c}
\includegraphics[width=1\linewidth]{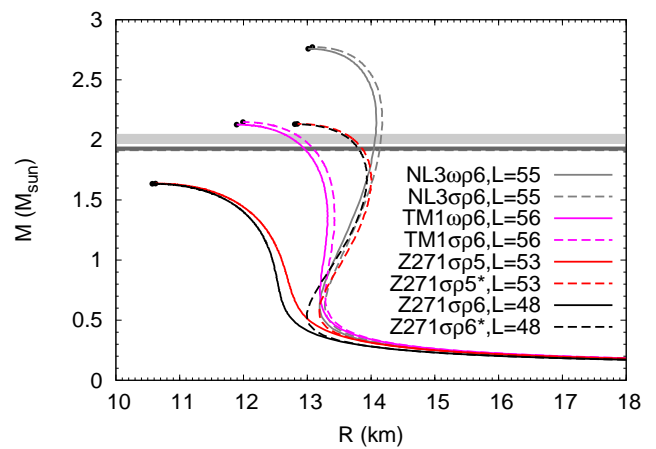}
  \end{tabular}
\caption{ Mass-radius relation of the models that passed almost all the constraints we considered. The black dots represent the maximum mass star. }
\label{fig11}
\end{figure}

We show in Figure \ref{fig11} the mass-radius relation for the models that passed almost all constraints we considered: the ones from Ref. \cite{Dutra-14}, the ones from Refs. \cite{Stone16, Stone14},  the microscopic neutron matter calculations \cite{Hebeler-13,Gandolfi12}, and the 2$M_{\odot}$ observational constraint \cite{Antoniadis13, Fonseca16}. We note that the Z271$\sigma\rho5-6$ models fail this last constraint, however using the procedure of Ref. \cite{Maslov-15}, we are able to get parametrizations Z271$\sigma\rho5*-6*$ that describe 2$M_{\odot}$ stars. Models NL3$\omega\rho6$,  NL3$\sigma\rho6$ and  Z271$\sigma\rho5*-6*$ fail the flow and KaoS experiments.
In Table \ref{tab8}, we show the properties of the 1.4$M_{\odot}$ stars obtained with those models, together with the transition density and pressure.  The following conclusions can be drawn: a) all models predict a similar transition density and have a similar symmetry energy slope; b) the $\sigma\rho$ models predict a lower transition pressure, $P_t\sim0.36\pm0.04$ MeVfm$^{-3}$, while for the $\omega\rho$ models the value is $P_t\sim0.5$MeVfm$^{-3}$; c) the radius and crust thickness of the   1.4$M_{\odot}$ stars within the models that predict 2$M_\odot$ stars are in the interval $13.3\lesssim R_{1.4}\lesssim13.9$ km and $1.3\lesssim \Delta R\lesssim 1.4$km, respectively.

\begin{table*}[htb]
\caption{ Some properties of the 1.4$M_\odot$ stars of the models that passed almost all the constraints, computed with an inner crust, using NL3$\sigma\rho$ or NL3$\omega\rho$.  The transition pressure, $P_t$, and density, $\rho_t$, to uniform matter are also shown.}  \label{tab8}
  \begin{tabular}{ccccccccccc}
    \hline
    \hline
 Model & \phantom{a} & $L$ (MeV) & \phantom{a} & $R_{1.4 M\odot}$ (km) &\phantom{a} & $\Delta R_{crust}$ (km) & \phantom{a} & $P_t$ (MeV/fm$^3$)& \phantom{a} & $\rho_t$ (fm$^{-3}$)\\
    \hline

NL3$\omega\rho$6 &\phantom{a}& 55 &\phantom{a}& 13.753 & \phantom{a} & 1.425 & \phantom{a} & 0.516 & \phantom{a} & 0.084 \\

NL3$\sigma\rho$6 &\phantom{a}& 55 &\phantom{a}& 13.846 & \phantom{a} & 1.400 & \phantom{a} & 0.382 & \phantom{a} & 0.081 \\

TM1$\omega\rho$6 &\phantom{a}& 56 &\phantom{a}& 13.317 & \phantom{a} & 1.323 & \phantom{a} & 0.497 & \phantom{a} & 0.082 \\

TM1$\sigma\rho$6 &\phantom{a}& 56 &\phantom{a}& 13.428 & \phantom{a} & 1.302 & \phantom{a} & 0.379 & \phantom{a} & 0.080 \\

Z271$\sigma\rho$5 &\phantom{a}& 53 &\phantom{a}& 12.110 &\phantom{a} & 1.035 & \phantom{a} & 0.398 & \phantom{a} & 0.083 \\

Z271$\sigma\rho5^*$ &\phantom{a}& 53 &\phantom{a}& 13.914 &\phantom{a} & 1.419 & \phantom{a} & 0.398 & \phantom{a} & 0.083 \\

Z271$\sigma\rho$6 &\phantom{a}& 48 &\phantom{a}& 12.001 & \phantom{a} & 0.995 & \phantom{a} &  0.323 & \phantom{a} & 0.085 \\

Z271$\sigma\rho6^*$ &\phantom{a}& 48 &\phantom{a}& 13.833 & \phantom{a} & 1.389 & \phantom{a} &  0.323 & \phantom{a} & 0.085 \\

    \hline
    \hline
  \end{tabular}
\end{table*}

\section{Conclusions} \label{IV}

In the present work we have generalized the Vlasov formalism developed in previous works  \cite{Avancini-05,ProvidenciaC-06a,Pais09,Pais10} with the $\omega$, $\rho$, $\sigma$-meson terms, by including mixed terms, up to fourth order. The dispersion relation obtained allows the study of the isoscalar collective modes of nuclear matter, and the instability modes that drive the system at subsaturation densities to a non-homogeneous phase. The dynamical spinodal surface is determined as the locus of the zero frequency isoscalar mode. The knowledge of the dynamical spinodal and the $\beta$-equilibrium EoS is used to make a good estimation of the crust-core transition density of a neutron star.

We have applied the formalism developed to study several families of stars that differ by a mixed $\omega\rho$ or $\sigma\rho$ term which was introduced to modify the density dependence of the symmetry energy. We have analysed the dependence of the crust-core transition density, $\rho_t$, and pressure, $P_t$, with the slope $L$ of the model. We have confirmed previous results, in particular, an almost linear anti-correlation between  $\rho_t$ and $L$, for both mixed  $\omega\rho$ or  $\sigma\rho$ terms. However, in a recent publication \cite{Pais16}, it has been shown that  a non-linear $\sigma\rho$ term of the form $\sigma\rho^2$, instead of $\sigma^2\rho^2$, as we have considered in the present work, did not show this linear behavior with $L$. The behavior of $P_t$ is non-monotonic, as discussed in \cite{ducoin10,Ducoin11}, taking the $\sigma\rho$ family.  However, if we consider only the $\omega\rho$  families, an anti-correlation of $P_t$ with $L$ is obtained. The largest transition pressures,  largest crust thicknesses and smallest radii   are obtained within the $\omega\rho$ families, see also \cite{Pais16}.  For the $\sigma\rho$ families, $P_t$ increases (decreases) with $L$, for $L<70$ ($L>70$) MeV.

It was shown how the knowledge of the crust-core transition density enables the construction of the stellar EoS and the determination of the mass-radius curve for the family of stars, within a given model with a small radius uncertainty. For the outer crust, that extends up to the neutron drip density,  the BPS EoS is considered. Between the neutron drip density and the crust-core transition density, the inner crust EoS should be taken. We have determined the inner crust EoS within a TF calculation for the $\omega\rho$ and $\sigma\rho$ NL3 family. It was shown that the error introduced in the calculation of the radius of a star with a mass above  $1 M_\odot$ is small, if  the inner crust of a model with similar isovector saturation properties is used to describe the inner crust.

 Since the Z21 family gives rise to a too soft EoS, a $\sigma$ dependent term was included in the Lagrangian density, as suggested in \cite{Maslov-15}. This extra term does not change the model properties at saturation density and below, but hardens the EoS above saturation density, allowing the description of 2 $M_\odot$, but, at the same time, not satisfying the KaoS restrictions. 

We propose a set of stellar matter EoS that satisfy well accepted saturation properties as well as constraints coming from microscopic neutron matter calculations, and from experimental results. All models predict a similar transition density of the order of $0.08$ fm$^{-3}$. However, there are differences in the transition pressure. The $\sigma\rho$ models predict a lower transition pressure than the $\omega\rho$ models. For the 1.4$M_\odot$ stars, these models predict, respectively,  a radius of $13.6\pm 0.3$ km and a crust thickness of $1.36\pm 0.06$km. These values for $R_{1.4}$ are  above the prediction of \cite{Lattimer14} but within the prediction of \cite{Fortin16}. 
 These crust thicknesses  are $\sim 25\%$ smaller than the one obtained in \cite{Piekarewicz-14} for the NL3max, which is close to our parametrization NL3$\omega\rho$5. We have considered the NL3max parametrization and we have obtained, using our formalism, $P_t=0.530$ MeV/fm$^3$ and $\rho_t=0.081$ fm$^{-3}$, just slightly below the corresponding quantities in \cite{Piekarewicz-14}, respectively, 0.550 MeV/fm$^3$ and 0.0826 fm$^{-3}$. However, for the  crust thickness, we have obtained  $\Delta R(M=1.4M_{\odot})=1.454$ km, that should be compared with $\Delta R(M=1.4M_{\odot})=1.990$ km in \cite{Piekarewicz-14}. The possible origin of this large difference is the EoS used in \cite{Piekarewicz-14} for the inner crust EoS, a politropic that matches the BPS EoS at the neutron drip density and the homogeneous EoS at the crust-core transition. As we have shown in the present work, a non adequate choice of the inner crust may introduce large uncertainties in the radius of low mass stars, see also discussion in \cite{Fortin16}.

\section*{ACKNOWLEDGMENTS}
 H.P. is supported by FCT under Project No. SFRH/BPD/95566/2013. Partial support comes from ``NewCompStar'', COST Action MP1304. The authors acknowledge the Laboratory for Advanced Computing at the University of Coimbra for providing CPU time with the Navigator cluster. We thank S. S. Avancini for providing the finite nuclei properties code.

\section{Appendix}

\newpage
\begin{longtable*}{c|ccccccccccccccccccc}
\caption{ Equation of state of the inner crust with pasta for the NL3$\omega\rho4$, NL3$\sigma\rho4$, and NL3$\sigma\rho6$ models. The energy density, $\varepsilon$, and pressure, $P$,  are in units of fm$^{-4}$. The pasta inner crust EoS for FSU and NL3$\omega\rho6$ are given in \cite{Grill-14}.}  \label{tab6} \\
\hline\hline \\[0.5pt]
             &   \multicolumn{2}{c}{NL3$\omega\rho4$} &\phantom{a} &\multicolumn{2}{c}{NL3$\sigma\rho4$} &\phantom{a} &\multicolumn{2}{c}{NL3$\sigma\rho6$} &\phantom{a} \\
$\rho$ (fm$^{-3}$)  &  \multicolumn{1}{c}{$\varepsilon$} & \multicolumn{1}{c}{P}  &\phantom{a} &  \multicolumn{1}{c}{$\varepsilon$} & \multicolumn{1}{c}{P} &\phantom{a} &  \multicolumn{1}{c}{$\varepsilon$} & \multicolumn{1}{c}{P}   \\[0.5pt]
\hline \\[0.5pt]
0.0020   & 0.009529489093     & 0.9527332622$\times$10$^{-5}$    &\phantom{a} & 0.009524920955        & 0.9071237400$\times$10$^{-5}$ &\phantom{a} & 0.009525820613 & 0.9831396710$\times 10^{-5}$ \\ [0.5pt]
0.0030   & 0.014300051145     & 0.1449370848$\times$10$^{-4}$    &\phantom{a} & 0.014293017797        & 0.1424032234$\times$10$^{-4}$ &\phantom{a} & 0.014294951223 & 0.1601402801$\times$10$^{-4}$ \\[0.5pt]
0.0040   & 0.019072361290     & 0.1981482455$\times$10$^{-4}$    &\phantom{a} & 0.019062936306        & 0.1981482455$\times$10$^{-4}$ &\phantom{a} & 0.019066335633 & 0.2321020474$\times$10$^{-4}$ \\[0.5pt]
0.0050   & 0.023846043274     & 0.2533865154$\times$10$^{-4}$    &\phantom{a} & 0.023834312335        & 0.2579474858$\times$10$^{-4}$ &\phantom{a} & 0.023839607835 & 0.3106518852$\times$10$^{-4}$ \\[0.5pt]
0.0060   & 0.028620865196     & 0.3111586557$\times$10$^{-4}$    &\phantom{a} & 0.028606912121        & 0.3197737897$\times$10$^{-4}$ &\phantom{a} & 0.028614535928 & 0.3962964911$\times$10$^{-4}$ \\[0.5pt]
0.0070   & 0.033396668732     & 0.3709578596$\times$10$^{-4}$    &\phantom{a} & 0.033380579203        & 0.3856542753$\times$10$^{-4}$ &\phantom{a} & 0.033390954137 & 0.4885291855$\times$10$^{-4}$ \\[0.5pt]
0.0080   & 0.038173336536     & 0.4322774112$\times$10$^{-4}$    &\phantom{a} & 0.038155212998        & 0.4540686496$\times$10$^{-4}$ &\phantom{a} & 0.038168746978 & 0.5878567026$\times$10$^{-4}$ \\[0.5pt]
0.0090   & 0.042950786650     & 0.4971443195$\times$10$^{-4}$    &\phantom{a} & 0.042930707335        & 0.5255236465$\times$10$^{-4}$ &\phantom{a} & 0.042947817594 & 0.6937723083$\times$10$^{-4}$ \\[0.5pt]
0.0100   & 0.047728978097     & 0.5640384188$\times$10$^{-4}$    &\phantom{a} & 0.047707032412        & 0.6015395775$\times$10$^{-4}$ &\phantom{a} & 0.047728102654 & 0.8062758570$\times$10$^{-4}$ \\[0.5pt]
0.0110   & 0.052507854998     & 0.6349865725$\times$10$^{-4}$    &\phantom{a} & 0.052484132349        & 0.6811029743$\times$10$^{-4}$ &\phantom{a} & 0.052509546280 & 0.9253675671$\times$10$^{-4}$ \\[0.5pt]
0.0120   & 0.057287395000     & 0.7104958058$\times$10$^{-4}$    &\phantom{a} & 0.057261981070        & 0.7652273052$\times$10$^{-4}$ &\phantom{a} & 0.057292107493 & 0.1050540523$\times$10$^{-3}$ \\[0.5pt]
0.0130   & 0.062067583203     & 0.7895524323$\times$10$^{-4}$    &\phantom{a} & 0.062040548772        & 0.8549261111$\times$10$^{-4}$ &\phantom{a} & 0.062075737864 & 0.1183315035$\times$10$^{-3}$ \\[0.5pt]
0.0140   & 0.066848404706     & 0.8746902313$\times$10$^{-4}$    &\phantom{a} & 0.066819831729        & 0.9501994646$\times$10$^{-4}$ &\phantom{a} & 0.066860415041 & 0.1322170865$\times$10$^{-3}$ \\[0.5pt]
0.0150   & 0.071629844606     & 0.9648958076$\times$10$^{-4}$    &\phantom{a} & 0.071599796414        & 0.1050540523$\times$10$^{-3}$ &\phantom{a} & 0.071646101773 & 0.1467614784$\times$10$^{-3}$ \\[0.5pt]
0.0160   & 0.076411917806     & 0.1062196243$\times$10$^{-3}$    &\phantom{a} & 0.076380468905        & 0.1157976367$\times$10$^{-3}$ &\phantom{a} & 0.076432794333 & 0.1619646646$\times$10$^{-3}$ \\[0.5pt]
0.0170   & 0.081194624305     & 0.1166591464$\times$10$^{-3}$    &\phantom{a} & 0.081161834300        & 0.1272507070$\times$10$^{-3}$ &\phantom{a} & 0.081220448017 & 0.1778773440$\times$10$^{-3}$ \\[0.5pt]
0.0180   & 0.085977941751     & 0.1277574775$\times$10$^{-3}$    &\phantom{a} & 0.085943877697        & 0.1392612321$\times$10$^{-3}$ &\phantom{a} & 0.086009055376 & 0.1943474635$\times$10$^{-3}$ \\[0.5pt]
0.0190   & 0.090761914849     & 0.1397173182$\times$10$^{-3}$    &\phantom{a} & 0.090726628900        & 0.1521839440$\times$10$^{-3}$ &\phantom{a} & 0.090798601508 & 0.2114763920$\times$10$^{-3}$ \\[0.5pt]
0.0200   & 0.095546536148     & 0.1524880063$\times$10$^{-3}$    &\phantom{a} & 0.095510080457        & 0.1659174886$\times$10$^{-3}$ &\phantom{a} & 0.095589056611 & 0.2292641148$\times$10$^{-3}$ \\[0.5pt]
0.0210   & 0.100331813097     & 0.1660695270$\times$10$^{-3}$    &\phantom{a} & 0.100294232368        & 0.1803605264$\times$10$^{-3}$ &\phantom{a} & 0.100380428135 & 0.2476093068$\times$10$^{-3}$ \\[0.5pt]
0.0220   & 0.105117775500     & 0.1806645887$\times$10$^{-3}$    &\phantom{a} & 0.105079092085        & 0.1957157365$\times$10$^{-3}$ &\phantom{a} & 0.105172678828 & 0.2665626234$\times$10$^{-3}$ \\[0.5pt]
0.0230   & 0.109904408455     & 0.1962225215$\times$10$^{-3}$    &\phantom{a} & 0.109864674509        & 0.2119831624$\times$10$^{-3}$ &\phantom{a} & 0.109965816140 & 0.2860733657$\times$10$^{-3}$ \\[0.5pt]
0.0240	 & 0.114691741765     &	0.2126419713$\times$10$^{-3}$    &\phantom{a} & 0.114650979638	 & 0.2290107223$\times$10$^{-3}$ &\phantom{a} & 0.114759802818 & 0.3061415919$\times$10$^{-3}$ \\[0.5pt]
0.0250	 & 0.119479775429     &	0.2302269859$\times$10$^{-3}$    &\phantom{a} & 0.119438014925	 & 0.2471531916$\times$10$^{-3}$ &\phantom{a} & 0.119554653764 & 0.3267672437$\times$10$^{-3}$ \\[0.5pt]
0.0260	 & 0.124268546700     &	0.2487241873$\times$10$^{-3}$    &\phantom{a} & 0.124225787818	 & 0.2661572071$\times$10$^{-3}$ &\phantom{a} & 0.124350324273 & 0.3478490107$\times$10$^{-3}$ \\[0.5pt]
0.0270	 & 0.129058048129     &	0.2684376668$\times$10$^{-3}$    &\phantom{a} & 0.129014313221	 & 0.2860226959$\times$10$^{-3}$ &\phantom{a} & 0.129146814346 & 0.3694882325$\times$10$^{-3}$ \\[0.5pt]
0.0280	 & 0.133848294616     &	0.2892660559$\times$10$^{-3}$    &\phantom{a} & 0.133803606033	 & 0.3070030943$\times$10$^{-3}$ &\phantom{a} & 0.133944109082 & 0.3915328707$\times$10$^{-3}$ \\[0.5pt]
0.0290	 & 0.138639286160     &	0.3110066173$\times$10$^{-3}$    &\phantom{a} & 0.138593643904	 & 0.3289463930$\times$10$^{-3}$ &\phantom{a} & 0.138742208481 & 0.4140842648$\times$10$^{-3}$ \\[0.5pt]
0.0300	 & 0.143431067467     &	0.3341154661$\times$10$^{-3}$    &\phantom{a} & 0.143384456635	 & 0.3517511941$\times$10$^{-3}$ &\phantom{a} & 0.143541097641 & 0.4371930845$\times$10$^{-3}$ \\[0.5pt]
0.0310	 & 0.148223638535     &	0.3583898942$\times$10$^{-3}$    &\phantom{a} & 0.148176059127	 & 0.3757722152$\times$10$^{-3}$ &\phantom{a} & 0.148340746760 & 0.4605046706$\times$10$^{-3}$ \\[0.5pt]
0.0320	 & 0.153017029166     &	0.3838805715$\times$10$^{-3}$    &\phantom{a} & 0.152968466282	 & 0.4008067772$\times$10$^{-3}$ &\phantom{a} & 0.153141170740 & 0.4840695765$\times$10$^{-3}$ \\[0.5pt]
0.0330	 & 0.157811194658     &	0.4102834500$\times$10$^{-3}$    &\phantom{a} & 0.157761648297	 & 0.4266015312$\times$10$^{-3}$ &\phantom{a} & 0.157942339778 & 0.5083946744$\times$10$^{-3}$ \\[0.5pt]
0.0340	 & 0.162606209517     &	0.4381559847$\times$10$^{-3}$    &\phantom{a} & 0.162555649877	 & 0.4537139030$\times$10$^{-3}$ &\phantom{a} & 0.162744253874 & 0.5328211701$\times$10$^{-3}$ \\[0.5pt]
0.0350	 & 0.167402043939     &	0.4672447394$\times$10$^{-3}$    &\phantom{a} & 0.167350441217	 & 0.4818398156$\times$10$^{-3}$ &\phantom{a} & 0.167546868324 & 0.5576530239$\times$10$^{-3}$ \\[0.5pt]
0.0360	 & 0.172198727727     &	0.4972963943$\times$10$^{-3}$    &\phantom{a} & 0.172146052122	 & 0.5110806087$\times$10$^{-3}$ &\phantom{a} & 0.172350198030 & 0.5827890127$\times$10$^{-3}$ \\[0.5pt]
0.0370	 & 0.176996275783     &	0.5288176471$\times$10$^{-3}$    &\phantom{a} & 0.176942497492	 & 0.5410308950$\times$10$^{-3}$ &\phantom{a} & 0.177154257894 & 0.6081783213$\times$10$^{-3}$ \\[0.5pt]
0.0380	 & 0.181794673204     &	0.5616059061$\times$10$^{-3}$    &\phantom{a} & 0.181739777327	 & 0.5723494687$\times$10$^{-3}$ &\phantom{a} & 0.181958958507 & 0.6339224055$\times$10$^{-3}$ \\[0.5pt]
0.0390	 & 0.186593979597     &	0.5956103560$\times$10$^{-3}$    &\phantom{a} & 0.186537876725	 & 0.6046815543$\times$10$^{-3}$ &\phantom{a} & 0.186764389277 & 0.6598691689$\times$10$^{-3}$ \\[0.5pt]
0.0400	 & 0.191394165158     &	0.6305776769$\times$10$^{-3}$    &\phantom{a} & 0.191336825490	 & 0.6381286075$\times$10$^{-3}$ &\phantom{a} & 0.191570460796 & 0.6861706497$\times$10$^{-3}$ \\[0.5pt]
0.0410	 & 0.196195229888     &	0.6671160227$\times$10$^{-3}$    &\phantom{a} & 0.196136608720	 & 0.6722850958$\times$10$^{-3}$ &\phantom{a} & 0.196377217770 & 0.7126749260$\times$10$^{-3}$ \\[0.5pt]
0.0420	 & 0.200997203588     &	0.7048706175$\times$10$^{-3}$    &\phantom{a} & 0.200937271118	 & 0.7078098715$\times$10$^{-3}$ &\phantom{a} & 0.201184600592 & 0.7393818232$\times$10$^{-3}$ \\[0.5pt]
0.0430	 & 0.205800086260     &	0.7435374428$\times$10$^{-3}$    &\phantom{a} & 0.205738782883	 & 0.7443482173$\times$10$^{-3}$ &\phantom{a} & 0.205992653966 & 0.7663421566$\times$10$^{-3}$ \\[0.5pt]
0.0440	 & 0.210603877902     &	0.7838258753$\times$10$^{-3}$    &\phantom{a} & 0.210541129112	 & 0.7819508319$\times$10$^{-3}$ &\phantom{a} & 0.210801303387 & 0.7935052272$\times$10$^{-3}$ \\[0.5pt]
0.0450	 & 0.215408638120     &	0.8252799162$\times$10$^{-3}$    &\phantom{a} & 0.215344354510	 & 0.8203641628$\times$10$^{-3}$ &\phantom{a} & 0.215610593557 & 0.8209216176$\times$10$^{-3}$ \\[0.5pt]
0.0460	 & 0.220214262605     &	0.8680008468$\times$10$^{-3}$    &\phantom{a} & 0.220148459077	 & 0.8600444999$\times$10$^{-3}$ &\phantom{a} & 0.220420479774 & 0.8487434825$\times$10$^{-3}$ \\[0.5pt]
0.0470	 & 0.225020855665     &	0.9116340661$\times$10$^{-3}$    &\phantom{a} & 0.224953413010	 & 0.9007891058$\times$10$^{-3}$ &\phantom{a} & 0.225230976939 & 0.8765652892$\times$10$^{-3}$ \\[0.5pt]
0.0480	 & 0.229828447104     &	0.9568381938$\times$10$^{-3}$    &\phantom{a} & 0.229759275913	 & 0.9425471653$\times$10$^{-3}$ &\phantom{a} & 0.230042085052 & 0.9045898332$\times$10$^{-3}$ \\[0.5pt]
0.0490	 & 0.234636932611     &	0.1003258629$\times$10$^{-2}$    &\phantom{a} & 0.234565988183	 & 0.9850654751$\times$10$^{-3}$ &\phantom{a} & 0.234853759408 & 0.9328678134$\times$10$^{-3}$ \\[0.5pt]
0.0500	 & 0.239446416497     &	0.1050844556$\times$10$^{-2}$    &\phantom{a} & 0.239373609424	 & 0.1028901315$\times$10$^{-2}$ &\phantom{a} & 0.239666014910 & 0.9612977155$\times$10$^{-3}$ \\[0.5pt]
0.0510   & 0.244256809354     & 0.1099342713$\times$10$^{-2}$    &\phantom{a} & 0.244182094932        & 0.1073801308$\times$10$^{-2}$ &\phantom{a} & 0.244478836656 & 0.9899811121$\times$10$^{-3}$ \\ [0.5pt]
0.0520   & 0.249068230391     & 0.1149259857$\times$10$^{-2}$    &\phantom{a} & 0.248991489410        & 0.1119664288$\times$10$^{-2}$ &\phantom{a} & 0.249292254448 & 0.1018968527$\times$10$^{-2}$ \\[0.5pt]
0.0530   & 0.253880590200     & 0.1200494706$\times$10$^{-2}$    &\phantom{a} & 0.253801763058        & 0.1166490139$\times$10$^{-2}$ &\phantom{a} & 0.254106193781 & 0.1048057340$\times$10$^{-2}$ \\[0.5pt]
0.0540   & 0.258693933487     & 0.1252894988$\times$10$^{-2}$    &\phantom{a} & 0.258612930775        & 0.1214076183$\times$10$^{-2}$ &\phantom{a} & 0.258920729160 & 0.1077298075$\times$10$^{-2}$ \\[0.5pt]
0.0550   & 0.263508260250     & 0.1306410180$\times$10$^{-2}$    &\phantom{a} & 0.263424992561        & 0.1263131737$\times$10$^{-2}$ &\phantom{a} & 0.263735800982 & 0.1106893644$\times$10$^{-2}$ \\[0.5pt]
0.0560   & 0.268323540688     & 0.1361090923$\times$10$^{-2}$    &\phantom{a} & 0.268237978220        & 0.1312998240$\times$10$^{-2}$ &\phantom{a} & 0.268551379442 & 0.1136742532$\times$10$^{-2}$ \\[0.5pt]
0.0570   & 0.273139834404     & 0.1416582614$\times$10$^{-2}$    &\phantom{a} & 0.273051828146        & 0.1363878255$\times$10$^{-2}$ &\phantom{a} & 0.273367494345 & 0.1166743576$\times$10$^{-2}$ \\[0.5pt]
0.0580   & 0.277957081795     & 0.1473594690$\times$10$^{-2}$    &\phantom{a} & 0.277866572142        & 0.1415771898$\times$10$^{-2}$ &\phantom{a} & 0.278184115887 & 0.1193450415$\times$10$^{-2}$ \\[0.5pt]
0.0590   & 0.282775372267     & 0.1531822840$\times$10$^{-2}$    &\phantom{a} & 0.282682240009        & 0.1468628179$\times$10$^{-2}$ &\phantom{a} & 0.283001244068 & 0.1224211650$\times$10$^{-2}$ \\[0.5pt]
0.0600   & 0.287594646215     & 0.1590811298$\times$10$^{-2}$    &\phantom{a} & 0.287498801947        & 0.1522396924$\times$10$^{-2}$ &\phantom{a} & 0.287818878889 & 0.1255124691$\times$10$^{-2}$ \\[0.5pt]
0.0610   & 0.292414903641     & 0.1651167870$\times$10$^{-2}$    &\phantom{a} & 0.292316287756        & 0.1577280345$\times$10$^{-2}$ &\phantom{a} & 0.292637050152 & 0.1286341925$\times$10$^{-2}$ \\[0.5pt]
0.0620   & 0.297236144543     & 0.1712639467$\times$10$^{-2}$    &\phantom{a} & 0.297134667635        & 0.1632873435$\times$10$^{-2}$ &\phantom{a} & 0.297455728054 & 0.1317913993$\times$10$^{-2}$ \\[0.5pt]
0.0630   & 0.302058428526     & 0.1775073935$\times$10$^{-2}$    &\phantom{a} & 0.301954001188        & 0.1689581317$\times$10$^{-2}$ &\phantom{a} & 0.302274912596 & 0.1349739265$\times$10$^{-2}$ \\[0.5pt]
0.0640   & 0.306881725788     & 0.1838724711$\times$10$^{-2}$    &\phantom{a} & 0.306774199009        & 0.1747201430$\times$10$^{-2}$ &\phantom{a} & 0.307094633579 & 0.1381919370$\times$10$^{-2}$ \\[0.5pt]
0.0650   & 0.311706036329     & 0.1902932650$\times$10$^{-2}$    &\phantom{a} & 0.311595320702        & 0.1802540966$\times$10$^{-2}$ &\phantom{a} & 0.311914891005 & 0.1414707513$\times$10$^{-2}$ \\[0.5pt]
0.0660   & 0.316531240940     & 0.1965164440$\times$10$^{-2}$    &\phantom{a} & 0.316417217255        & 0.1862238860$\times$10$^{-2}$ &\phantom{a} & 0.316735535860 & 0.1447597286$\times$10$^{-2}$ \\[0.5pt]
0.0670   & 0.321357458830     & 0.2031500917$\times$10$^{-2}$    &\phantom{a} & 0.321240127087        & 0.1922595431$\times$10$^{-2}$ &\phantom{a} & 0.321556776762 & 0.1480993466$\times$10$^{-2}$ \\[0.5pt]
0.0680   & 0.326184719801     & 0.2099307254$\times$10$^{-2}$    &\phantom{a} & 0.326063960791        & 0.1983914990$\times$10$^{-2}$ &\phantom{a} & 0.326378494501 & 0.1514491159$\times$10$^{-2}$ \\[0.5pt]
0.0690   & 0.331013083458     & 0.2167164115$\times$10$^{-2}$    &\phantom{a} & 0.330888777971        & 0.2045741305$\times$10$^{-2}$ &\phantom{a} & 0.331200808287 & 0.1548850443$\times$10$^{-2}$ \\[0.5pt]
0.0700   & 0.335842400789     & 0.2236136002$\times$10$^{-2}$    &\phantom{a} & 0.335714370012        & 0.2098242985$\times$10$^{-2}$ &\phantom{a} & 0.336023390293 & 0.1571097760$\times$10$^{-2}$ \\[0.5pt]
0.0710   & 0.340671092272     & 0.2294718986$\times$10$^{-2}$    &\phantom{a} & 0.340540796518        & 0.2163312631$\times$10$^{-2}$ &\phantom{a} & 0.340846419334 & 0.1606267877$\times$10$^{-2}$ \\[0.5pt]
0.0720   & 0.345502316952     & 0.2361764899$\times$10$^{-2}$    &\phantom{a} & 0.345368176699        & 0.2228483791$\times$10$^{-2}$ &\phantom{a} & 0.345669955015 & 0.1643870375$\times$10$^{-2}$ \\[0.5pt]
0.0730   &     		      &     		    &\phantom{a} &         		 &		    &\phantom{a} & 0.350494056940 & 0.1680915593$\times$10$^{-2}$ \\[0.5pt]
0.0740	 &		      &     		    &\phantom{a} &	     		 &		    &\phantom{a} & 0.355318605900 & 0.1718517975$\times$10$^{-2}$ \\[0.5pt]
0.0750	 &		      &     		    &\phantom{a} &	     		 &		    &\phantom{a} & 0.360143750906 & 0.1756627345$\times$10$^{-2}$ \\[0.5pt]
0.0760	 &		      &     		    &\phantom{a} &	     		 &		    &\phantom{a} & 0.364969402552 & 0.1795750228$\times$10$^{-2}$ \\[0.5pt]
0.0770	 &		      &     		    &\phantom{a} &	     		 &		    &\phantom{a} & 0.369795531034 & 0.1834822469$\times$10$^{-2}$ \\[0.5pt]
0.0780	 &		      &     		    &\phantom{a} &	     		 &		    &\phantom{a} & 0.374622225761 & 0.1874604146$\times$10$^{-2}$ \\[0.5pt]
0.0790	 &		      &     		    &\phantom{a} &	     		 &		    &\phantom{a} & 0.379449486732 & 0.1914335182$\times$10$^{-2}$ \\[0.5pt]
0.0800	 &		      &     		    &\phantom{a} &	     		 &		    &\phantom{a} & 0.384277373552 & 0.1953660743$\times$10$^{-2}$ \\[0.5pt]

    \hline
    \hline
\end{longtable*}

\end{document}